\documentclass[smallextended,runningheads]{svjour3}
\usepackage{amsmath,amssymb,amsfonts,amsbsy}

\usepackage[T1]{fontenc}
\usepackage{enumerate,url}

\usepackage{mathtools}
\usepackage[hidelinks]{hyperref}
\usepackage[nodoi]{apacite}
\usepackage{natbib}
\setcitestyle{round,aysep={}}
\usepackage[british]{babel}

\usepackage{xcolor}

\usepackage[utf8]{inputenc} 	
\usepackage[ruled]{algorithm2e}
\usepackage{comment}
\usepackage{tikz}
\usetikzlibrary{arrows.meta}
\usetikzlibrary{shapes.geometric}
\usepackage[normalem]{ulem}
 \usepackage{mathptmx}      
\usepackage{latexsym}

\newcommand{\hmi}{\hat{m}_1}
\newcommand{\hmo}{\hat{m}_0}

\newcommand{\tmi}{\tilde{m}_1}
\newcommand{\probfm}[1][m_1]{\mathbb{P}_{#1}}
\newcommand{\BH}[1][opt]{BH_{#1}}
\newcommand{\BY}[1][opt]{BY_{#1}}
\newcommand{\Exp}[1][opt]{Exp_{#1}}
\newcommand{\AORC}[1][opt]{AORC_{#1}}

\newcommand{\tr}{r}

\newcommand{\gstar}[1][\tmi]{\gamma_{#1}}

\newcommand{\tval}{Q}
\newcommand{\tobs}{q}

\newtheorem{defi}{Definition}
\newtheorem{prop}{Proposition}
\newtheorem{assumption}{Assumption}

\smartqed  
\usepackage{graphicx}

\begin{document}
\title{Confidence bounds for the true discovery proportion based on the exact distribution of the number of rejections
}

\titlerunning{Confidence bounds for the true discovery proportion} 

\author{Friederike Preusse \and
        Anna Vesely \and
        Thorsten Dickhaus
}

\authorrunning{F. Preusse \and A. Vesely \and T. Dickhaus} 

\institute{F. Preusse \at
            \email{preusse@uni-bremen.de}\\
            Institute for Statistics, University of Bremen,  D-28344 Bremen, Germany \and
           A. Vesely \at
            \email{vesely@uni-bremen.de}\\
            Institute for Statistics, University of Bremen,  D-28344 Bremen, Germany \and
            T. Dickhaus \at
            \email{dickhaus@uni-bremen.de}
            \\
            Institute for Statistics, University of Bremen,  D-28344 Bremen, Germany 
}

\date\today

\maketitle
\begin{abstract} 
In multiple hypotheses testing it has become widely popular to make inference on the true discovery proportion (TDP) of a set $\mathcal{M}$ of null hypotheses. This approach is useful for several application fields, such as neuroimaging and genomics. Several procedures to compute simultaneous lower confidence bounds for the TDP have been suggested in prior literature. Simultaneity allows for post-hoc selection of $\mathcal{M}$. If sets of interest are specified a priori, it is possible to gain power by removing the simultaneity requirement. We present an approach to compute lower confidence bounds for the TDP if the set of null hypotheses is defined a priori. The proposed method determines the bounds using the exact distribution of the number of rejections based on a step-up multiple testing procedure under independence assumptions. We assess robustness properties of our procedure and apply it to real data from the field of functional magnetic resonance imaging.

\end{abstract}
\newpage

\section{Introduction} \label{sec:intro}
When testing multiple null hypotheses simultaneously it can be of interest to make inference on sets of null hypotheses instead of on every single null hypothesis. Indeed, inference on each individual null hypothesis with appropriate error control may be very conservative, especially if the number of null hypotheses is large. Moreover, such a fine-grained analysis may not be of particular interest in practical applications. For example, in neuroimaging, brain activation in response to a stimulus is measured at the level of small-scale volume units known as voxels, but the primary interest typically lies in making inference on brain regions that comprise multiple voxels. 
Similarly, in genomics, inference is often made on collections of genes. 
In this setting, a global test on a set of null hypotheses only infers whether the set contains at least one false null hypothesis, i.e., at least one true discovery. However, if the set is large, knowing that there is at least one true discovery is unspecific in terms of the total number of true discoveries in the set. 

Therefore, in many cases it is of interest to make inference on the number of true discoveries or the corresponding proportion over all the considered null hypotheses (true discovery proportion, TDP). In particular, the computation of lower confidence bounds for the TDP is an active topic of research, \citep[see, e.g.,][]{Katsevich2020,Blain2022,Andreella2023,Tian2023,Vesely2023}. These bounds can be used to answer the following question: If we were to reject all null hypotheses within a set of null hypotheses of interest, how many true discoveries would we find? \citep{GoemanSolari}. 

Many procedures that define TDP confidence bounds, including the aforementioned methods, provide them simultaneously for all possible subsets of null hypotheses. This allows the researcher to choose the set of null hypotheses of interest post-hoc, that is, based on the computed confidence bound of the TDP. This is helpful in exploratory research when the sets of null hypotheses of interest have not been defined beforehand. However, the simultaneity requirement potentially reduces power (in a given sense, which we will explain formally in Section \ref{sec:critical_vector}) and can be removed when the set of interest is known a priori. For instance, in neuroimaging, sometimes researchers may want to explore different brain regions, but in other cases a region of interest is defined a priori (e.g., to test the reproducibility of a study, or to examine whether a region that is known to be associated with a certain stimulus is also associated with a similar one).

In this paper we propose a method to compute lower confidence bounds for the TDP if the subset of interest is specified a priori, assuming that the p-values are independent and their distribution under the alternative is known.
The bounds are derived from the exact distribution of the number of rejections of a step-up multiple testing procedure. The proposed method is flexible as any critical vector can be utilized for the step-up procedure. Furthermore, the choice of the critical vector influences the power of the proposed method. Therefore, we derive guidelines on how to select a suitable critical vector. We compare through simulations our proposal to the method of \citet{GoemanSolari}, on which most of other proposals in literature are based, demonstrating that we gain power in some settings, especially for low signal strength. Then, we evaluate the robustness of the proposed procedure against violations of the assumptions, in particular positive dependency between the p-values and misspecification of their distribution under the alternative.

The paper is structured as follows. We review established methods for the computation of lower confidence bounds for the TDP and briefly discuss step-up procedures in Section 2. In Section 3, we introduce the assumptions and our proposed methodology. 
The choice of the critical vector of the step-up procedure is discussed in Section 4. We compare the performance of the proposed methodology to the procedure by \citet{GoemanSolari} in a simulation study in Section 5 and investigate under which conditions the procedure still yields valid confidence bounds when the model assumptions are violated. In Section 6 we apply the method to a real fMRI data set. We give a brief summary and conclusion in Section 7. Proofs and additional results are given in the Appendix.

\section{Background and related literature}
\label{sec:Background}

In this section we introduce the framework and notation, and briefly discuss related work. First we consider methods to make inference on the TDP, reviewing existing literature; subsequently we introduce step-up tests on which our proposed method will rely. 

\subsection{Inference on the TDP} 

Let $\mathcal{M}=\{H_1,\ldots,H_m\}$ be a non-empty collection of null hypotheses and denote by $\mathcal{M}_1\subseteq \mathcal{M}$ the unknown subset of false null hypotheses (true discoveries). We are interested in making inference on the number of true discoveries $m_1=|\mathcal{M}_1|$, where $|\cdot|$ denotes the size of a set, or equivalently the proportion of true discoveries $\pi_1=m_1/m$. In fMRI data analysis, for instance, $\mathcal{M}$ may correspond to a region of the brain having $m$ voxels, while $m_1$ and $\pi_1$ are the number and the proportion of truly active voxels (which equals the TDP in that context), respectively.

For any $\alpha\in (0,1)$, we will compute a lower $(1-\alpha)$-confidence bound for the number of true discoveries, i.e., a value $\hmi$ such that
\begin{equation}
\mathbb{P}(m_1\geq \hmi)\geq 1-\alpha,\label{eq:def_bound}
\end{equation}
where $\mathbb{P}$ refers to the probability measure under the true but unknown data generating process. 
From $\hmi$ we can immediately derive confidence bounds for other quantities of interest such as the TDP and the number or proportion of false discoveries \citep{GoemanSolari}. For instance, a lower confidence bound for the TDP is $\hat{\pi}_1=\hmi/m$, while an upper confidence bound for the number of false discoveries is $\hmo=m-\hmi$.

Many procedures have been proposed that focus on simultaneous inference on the TDP based on confidence bounds \citep{GenoveseWasserman2006,Meinshausen2006,GoemanSolari,Lee2012,Rosenblatt,Hemerik2019,Blanchard2020,Ebrahimpoor2020,Blain2022,Cai2022,Andreella2023,Vesely2023}.
Such procedures make inference on the TDP simultaneously over all possible subsets of hypotheses. In this case, the set $\mathcal{M}$ is allowed to vary as any subset of a bigger multiple testing problem. For example, in fMRI data, different brain regions can be analyzed simultaneously.
This allows for post-hoc inference, as the confidence bounds are valid even if we select the subset of interest after computing the bounds.
\citet{GoemanSolari} emphasize this exploratory aspect of simultaneous TDP confidence bounds and show that simultaneity of the bounds is assured by embedding any global test into the closed testing framework \citep{Marcus1976}. It has been shown that this approach is equivalent to an earlier proposal by \citet{GenoveseWasserman2006} \citep{Goeman2021} and, if permutation tests are used, these bounds are identical to those proposed in \citet{Meinshausen2006} \citep{GoemanSolari}. 
First applications of the method to fMRI data \citep{Rosenblatt} and to genomics data \citep{Ebrahimpoor2020} rely on the comparison between a vector of ordered p-values and the critical vector based on the Simes test \citep{Simes1986}. A more flexible variant of the method is proposed in \citet{Hemerik2019} and \citet{Andreella2023}, who use conditional resampling to define a critical vector. To do so, families of critical vectors, i.e., collections of critical vectors based on the same critical value function, are defined a priori. Randomization is used to define the critical vector from the family, such that the chosen critical vector provides the highest valid confidence bound for the TDP. 
Similar procedures are proposed by \citet{Blanchard2020} and \citet{Blain2022}. The latter extends the method by constructing the families of critical vectors using additional randomization. Simultaneous inference on the TDP is an active topic of research, further developments can be found in \citet{Katsevich2020}, \citet{Davenport2022}, \citet{Chen2022}, \citet{Tian2023} and \citet{Goeman2023}.

If the set of null hypotheses $\mathcal{M}$ that we want to make inference on is selected a priori, we can potentially gain power by removing the simultaneity requirement. Such a confidence bound for the TDP has been proposed in \citet{Patra2016} for the two-groups mixture model of the p-values, as defined by \citet{Efron2001}. Our proposed method assumes a conditional version of the two-group mixture model, in accordance with \citet{RoquainVillers}. 
Similarly to \citet{Hemerik2019}, \citet{Andreella2023} and \citet{Blanchard2020}, our procedure relies on a critical vector chosen from an a priori defined family of critical vectors. Contrarily to their methods, all critical vectors in the family provide valid confidence bounds, so we can choose the critical vector that provides the highest bound possible.
We assume that the distribution of the p-values is known, but demonstrate through simulations that the method is robust to certain misspecifications of their distribution.

\subsection{Step-up procedures}
In this section we will give a short introduction to step-up tests. We first define the properties of such a procedure and then present some established tests. An exhaustive introduction to the topic is given in \citet{dickhaus_2014}, Chapter 5.

Multiple testing procedures are defined as functions that return a set of rejected null hypotheses based on the observed p-values. Formally, let $P_1,\ldots,P_m$ be random p-values for the null hypotheses of interest, 
and let $p_1,\ldots,p_m$ be their observed values. A multiple testing procedure rejects a subset $\mathcal{R}\subseteq\mathcal{M}$ of null hypotheses.
A step-up test is a particular multiple testing procedure that determines $\mathcal{R}$ by sorting the observed p-values and comparing them to a critical vector with the following properties.
\begin{defi}
\label{def:rejection_curves}
    A vector $(t_1,\ldots,t_m)^\top\in [0,1]^{m}$ is a critical vector for a step-up procedure if it is non-decreasing: $0\leq t_1 \leq \ldots\leq t_m \leq 1$. 
\end{defi}
Given a critical vector, the step-up procedure is defined as follows.
\begin{defi} \label{def:stepup}
     Denote by $(p_{1:m},\ldots,p_{m:m})^\top\in [0,1]^{m}$ the vector of ordered observed p-values, so that $p_{1:m}\leq\ldots\leq p_{m:m}$. Let $H_{1:m},\ldots, H_{m:m}$ be the corresponding ordered null hypotheses. Then let $k= max\{i=1,\ldots,m\,|\,p_{i:m}\leq t_i)$. A step-up procedure rejects $\mathcal{R}=\{H_{1:m},\ldots, H_{k:m}\}$ if $k$ exists, and $\mathcal{R}=\emptyset$ otherwise.
\end{defi}

Step-up tests can be constructed to control the family-wise error rate (FWER), the false discovery rate (FDR), or other error measures. Critical vectors change based on the desired error rate control level. While our method makes use of a generic step-up test, without requiring control of the FWER or FDR, we use the structure of established tests (i.e., established families of critical vectors) to derive critical vectors. In particular, we will focus on critical vectors defined for FDR control.
The most famous procedure is the linear step-up test by \cite{BenjaminiHochberg}, which is based on the critical value function by \cite{Simes1986}.
The power of the test can be improved by including additional information in the procedure, such as an estimate for the number of true null hypotheses \citep[for an overview and comparison]{BenjaminiHochberg2000,Sarkar2002,Storey2004,Hwang2011}. \citet{Finner2009} studied the asymptotic behavior of step-up-down tests (i.e., $m\to\infty$) and proposed an asymptotically optimal rejection curve (AORC), with optimality in the sense of optimal power. Critical value functions based on the AORC are given in \citet{Finner2009}, \citet{Gontscharuk2010}, \citet{Finneretal2012} and \citet{Habiger2014}. For arbitrary dependency among the p-values, \cite{BenjaminiYekutieli} adapted the procedure by \cite{BenjaminiHochberg}, while a new family of step-up tests is proposed in \citet{Blanchard2008}.

Properties of step-up tests are investigated in \citet{FinnerRoters2002}, \citet{Cohen2005}, \citet{FerreiraZwinderman2006}, \citet{RoquainVillers} and \citet{Chenetal2011}, among others. Under the assumption that all null hypotheses are true, \citet{FinnerRoters2002} derived the exact distribution of the number of rejected hypotheses for step-up procedures.
The exact distribution of the false discovery proportion for step-up and step-down tests for different p-value models is given in \citet{RoquainVillers}.
Explicit formulas for different notions of power for step-wise procedures are given in \citet{Chenetal2011}. Scenarios in which step-up tests cannot be employed are described in \citet{Cohen2005}.

\section{Methodology}
\label{sec:methodology}

In this section we propose a procedure to obtain a lower confidence bound for the number $m_1$ of true discoveries, denoted by $\hat{m}_1$ as in Eq.~\eqref{eq:def_bound}.
We assume that the p-value follow the model referred to as ``unconditional independent model'' by \citet{RoquainVillers}, so that the stochastically independent p-values corresponding to false null hypotheses follow the same cumulative distribution function (CDF) $F$.
\begin{assumption} \label{A:model}
The p-values $P_1,\ldots,P_m$ are stochastically independent and have the following (marginal) distribution:
\begin{equation*}
    P_i \sim \begin{cases}
        F &\text{if }1\leq i\leq m_1\\
        Uni[0,1] &\text{if }m_1+1\leq i\leq m.
    \end{cases}
\end{equation*}
\end{assumption}

While $F$ is assumed to be known, the number of true discoveries $m_1$ is not. We denote by $\probfm$ the probability measure under this model and compute the desired confidence bound in Eq.~\eqref{eq:def_bound} under this probability measure. For simplicity of notation, we notationally omit dependency on $m$ and $F$, as well as any other fixed quantities defined throughout the paper.

As mentioned in the previous section, we rely on a step-up procedure. Hence consider a step-up procedure, as given in Definition \ref{def:stepup}, that uses a pre-fixed critical vector $(t_1,\ldots,t_m)^\top$ and rejects a subset $\mathcal{R}$ of hypotheses. Denote by $R=|\mathcal{R}|$ the random variable representing the number of rejections and denote by $r$ its observed value. Note that $R$ is a function of the p-values that takes values in $\{0,\ldots,m\}$. 

Fix any $\alpha\in (0,1)$. The following theorem shows how to compute the desired lower ($1-\alpha$)-confidence bound $\hat{m_1}$, following from Remark 2 in \citet{vonSchroeder}. Throughout the paper, we denote by $\tilde{m}_1\in \{0,\ldots,m\}$ a generic candidate for the number of true discoveries, and by $\probfm[\tmi]$ the probability measure under the same model of Assumption \ref{A:model}, but having $\tmi$ false null hypotheses instead of $m_1$.

\begin{theorem}
\label{theorem:lower_bound}
For any $\tmi\in\{0,\ldots,m\}$, let
\begin{equation}
\gamma_{\tmi} =\max\{\gamma\in [0,1]\;|\;\probfm[\tmi](\tmi\geq R\gamma)\geq 1-\alpha\}. \label{eq:gammatmi}
\end{equation}
Subsequently, define
\begin{equation}
\hmi=\lceil r\gamma^*\rceil \qquad\text{as well as}\qquad \gamma^*= \min_{\tmi\in\{0,\ldots,m\}} \gamma_{\tmi},  \label{eq:hmi}
\end{equation}
where $\lceil\cdot\rceil$ represents the ceiling function. Then
\begin{equation}
\probfm(m_1\geq \hmi)\geq 1-\alpha. \label{eq:def_bound2}
\end{equation}
\end{theorem} 

A formal proof is provided in Appendix \ref{sec:allproofs}.
The main idea is to consider each possible value $\tmi$ for the number of true discoveries, ranging from $0$ to $m$. Under the corresponding model and probability measure $\probfm[\tmi]$, we determine $\gamma_{\tmi}$ as the largest value such that $\probfm[\tmi](\tmi\geq  R\gamma_{\tmi})\geq 1-\alpha$. Notice that here all quantities are fixed, with the exception of the random variable $R$. Subsequently, we take $\gamma^*$ as the minimum of all values $\gamma_{\tmi}$. It follows immediately that $\probfm[\tmi](\tmi\geq  R\gamma^*)\geq 1-\alpha$ for any $\tmi$, and in particular this holds for the true, unknown value $m_1$. Finally, the use of the ceiling function follows from the fact that $m_1$ can take only integer values.

From $\hat{m}_1$ we can immediately derive a confidence bound for the TDP by using $\hat{\pi}_1=\hmi/m$. We underline that the number of true discoveries and the TDP are defined considering the entire set $\mathcal{M}$ of hypotheses, and not with respect to the subset $\mathcal{R}$ of hypotheses that are rejected by the step-up procedure. Indeed, the step-up procedure is used only as a tool to construct the desired confidence bound, but the procedure does not require to associate any error control with $\mathcal{R}$. In the next sections we will show how to compute the confidence bound $\hmi$ in practice.

\subsection{Computation of the lower confidence bound}
To compute $\gamma^*$ and thus the desired confidence bound $\hmi$ we need to determine $\gamma_{\tmi}$ as in Eq.~\eqref{eq:gammatmi} for all possible values of $\tmi$ and then to evaluate the minimum.

Fix any $\tmi\in\{0,\ldots,m\}$. Eq.~\eqref{eq:gammatmi} requires to explore all values $\gamma\in [0,1]$. The following theorem shows that $\gamma_{\tmi}$ can be computed using the distribution of the number of rejections $R$ under $\probfm[\tmi]$. 

\begin{theorem}
\label{theorem:compute_gamma}
For $\tmi=0$,
\begin{equation*}
\gamma_{0}=\begin{cases}
        1&\text{ if $\probfm[\tmi](R=0)\geq 1-\alpha$}\\
        0 &\text{ otherwise.}
    \end{cases}
\end{equation*}
For $\tmi\in \{1,\ldots,m\}$, $\gamma_{\tmi} = \tmi / \ell_{\tmi}$, 
where
\begin{equation*}
    \ell_{\tmi}=\min\left\{\ell\in\{\tmi,\ldots,m-1,m\}\,:\,\probfm[\tmi](R\leq \ell)\geq 1-\alpha\right\}.
\end{equation*}
\end{theorem}
As $R$ is a discrete variable, Theorem \ref{theorem:compute_gamma} shows that $\gamma_{\tmi}$ can be computed evaluating the probability distribution of $R$ on a finite number of points, namely one if $\tmi=0$ and $m-\tmi$ otherwise. Furthermore, note that $\gamma_{\tmi}$ and therefore $\gamma^*$ do not depend on observations and do not change as long as the distribution of $R$ does not change. This allows us to compute $\gamma^*$ as soon as $F$ is known, before seeing the data, since $R$ is a function of the p-values. The observed data in form of the observed number of rejected hypotheses $r$ are only used when computing $\hmi$ from Eq.~\eqref{eq:hmi}.

In the next section we give the distribution of $R$ for step-up tests.

\subsection{Distribution of the number of rejections \texorpdfstring{$R$}{R}}

Equations given in \cite{RoquainVillers}, Section 5.3, can be used to compute the distribution of the number of rejections $R$. As in the previous sections, consider a generic step-up test that relies on a critical vector $(t_{1},\ldots,t_{m})^\top$. The following proposition gives the distribution of the random number $R$ of rejections of this step-up test.
\begin{prop}
    \label{prop:distribution_R}
    Fix any $\tmi\in\{0,\ldots,m\}$ and the probability measure $\probfm[\tmi]$ under the corresponding model. Then, for any $\ell\in\{0,\ldots,m\}$,
    \begin{align*} 
    \probfm[\tmi](R \leq \ell)&=\sum_{k=0}^{\ell}\sum_{j=0}^k\binom{m-\tmi}{j}\binom{\tmi}{k-j}t_k^j(F(t_k))^{k-j}\\
    &\quad \cdot\Psi_{m-\tmi-j,\tmi-k+j}^{Uni[0,1], \overline{F}}(1-t_m,\dots,1-t_{k+1}),\notag
\end{align*}
where $\overline{F}(t)=1-F(1-t)$ and $\Psi_{m-\tmi-j,\tmi-k+j}^{Uni[0,1], \overline{F}}$ denotes the joint distribution of $m-k$ ordered independent random variables, where $m-\tmi-j$ follow the uniform distribution $Uni[0,1]$ and the remaining ones have CDF $F$. This relates to the joint distribution of the ordered p-values that correspond to hypotheses not rejected by the underlying step-up procedure.
\end{prop} 

The joint distribution of order statistics can be computed using generalized recursions as given in \cite{vonSchroeder}.

In summary, the lower $(1-\alpha)$-confidence bound $\hat{m}_1$ for the number of true discoveries given in Theorem \ref{theorem:lower_bound} is computed considering all possible values $\tmi\in[0,m]$, evaluating for each of them the maximum of a given function defined from the distribution of the random number of rejections $R$ of a step-up test. This maximum can be determined by computing the distribution of $R$ for a finite number of points.
The algorithm to compute $\hat{m}_1$ is given in Appendix \ref{sec:Algorithm}.

\section{Choice of the critical vector}
\label{sec:critical_vector}
Any choice of the critical vector of the step-up procedure that satisfies Definition \ref{def:rejection_curves} leads to a valid confidence bound $\hmi$; however, this choice influences the procedure's performance. 
In this section, we present some of the families of critical vectors discussed in Section \ref{sec:Background} as well as a new one, then we illustrate how to select a suitable critical vector from a family which is assumed to be fixed a priori.
Suggestions on the choice of the family of critical vectors are given in Section \ref{sec:Simulation}.

A family of critical vectors can be defined as a collection of curves depending on one or more parameters. A particular critical vector $(t_1,\ldots,t_m)^\top$ can be chosen from the family by fixing the parameter values. A first example is the family of linear critical vectors proposed by \citet{BenjaminiHochberg}, defined by
\begin{equation}
    \label{eq:BH}
    t_i(\lambda)=\frac{i}{m}\lambda, \qquad \lambda\in [0,1].
\end{equation}
We denote this family by ``$BH$''. Another linear approach is that of \citet{BenjaminiYekutieli}:
\begin{equation}
    \label{eq:BY}
    t_i(\lambda)=\frac{i/m\cdot\lambda}{\sum\nolimits_{j=1}^m\frac{1}{j}}, \qquad \lambda\in \left[0,\sum\limits_{j=1}^m \frac{1}{j}\right]
\end{equation}
which we denote by ``$BY$''.
Critical vectors based on the AORC can also be employed. One possible family is given by \citet{Finner2009}:
\begin{equation}
\label{eq:AORC1}
    t_i(\lambda, \beta)=\frac{i\cdot\lambda}{m+\beta-i\cdot(1-\lambda)},\qquad \lambda \geq0,\quad \beta \geq 0.
\end{equation}
This family is denoted by ``$AORC$''. Note that $\lambda$ could also be negative, then $\beta < -m$, however, we focus on $\lambda\leq0$ due to its original interpretation as the level of FDR control in \citet{Finner2009}. Furthermore, we consider a new family which has a more flexible shape than $AORC$, $BH$ and $BY$: 
\begin{align}
    \label{eq:Exp2}
    t_i(\lambda, \beta) &= \lambda\cdot\left(\frac{i}{m}\right)^\beta, \qquad 0\leq \lambda \leq1,\quad \beta \geq 0. 
    \end{align} 
Depending on $\beta$ it is either a convex function ($\beta>1$), or a concave function ($\beta<1$), or a linear function ($\beta = 1$). This procedure is denoted by ``$Exp$''.

Note that the choice of the parameters, and consequently of the critical vector used in the procedure, is up to the practitioner. 
However, it is desirable that the procedure has high power and low variability, i.e., that the resulting confidence bound has a high expected value $E_{m_1}[\hmi]$ and low variance $Var_{m_1}[\hmi]$ under the true data generating model. In the following paragraphs we illustrate how to select the parameters in order to meet these requirements.

Fix any critical vector $(t_1,\ldots,t_m)^\top$, for which the corresponding step-up procedure rejects $R$ hypotheses. Recall that $\hmi=\lceil r\gamma^*\rceil$ depends on the realization of the random variable $R$ and on $\gamma^*$, which in turn depends on the distribution of $R$. Since such distribution is known, it is possible to determine the expected value and variance of $\hmi$ with respect to any probability measure $\probfm[\tmi]$. However, these values cannot be computed in closed form due to the ceiling function appearing in Eq.~\eqref{eq:hmi}; for this reason, we consider $r\gamma^*\approx \hmi$ instead.

\begin{prop}
 Fix any $\tmi\in\{0,\ldots,m\}$ and the probability measure $\probfm[\tmi]$ under the corresponding model. Then 
\label{Prob:Var_Mean}
    \begin{align*}
    E_{\tmi}[R\gamma^*]& = \gamma^*\cdot E_{\tmi}[R]\notag\\
    &=\gamma^* \cdot\sum\limits_{\ell=0}^m \ell\cdot\Big(\sum\limits_{j=0}^\ell\binom{m-\tmi}{j}\binom{\tmi}{\ell-j}(t_\ell)^j(F(t_\ell))^{\ell -j}\\
    &\quad\cdot\Psi_{m-\tmi-j, \tmi-\ell +j}^{Uni[0,1], F}(1-t_m,\ldots,1-t_{\ell +1})\Big)\notag.
\end{align*}
Furthermore, 
\begin{equation*}
    Var_{\tmi}[R\gamma^*]= (\gamma^*)^2\cdot Var_{\tmi}[R]
\end{equation*}
with
\begin{alignat*}{2}
        Var_{\tmi}[R]=&\sum_{\ell=0}^m \ell^2\cdot\Big[\sum_{j=0}^\ell&&\binom{m-\tmi}{j}\binom{\tmi}{\ell-j}(t_\ell)^j F(t_\ell)^{\ell-j}\notag \\
        & &&\cdot\Psi_{m-\tmi-j, \tmi-\ell+j}^{Uni[0,1], \Bar{F}}(1-t_m,\ldots,1-t_{\ell+1})\Big]\\
        &-\Big(\sum_{\ell=0}^m \ell\cdot\Big[\sum_{j=0}^\ell&&\binom{m-\tmi}{j}\binom{\tmi}{\ell-j}(t_\ell)^j F(t_\ell)^{\ell-j} \notag\\
        & &&\cdot\Psi_{m-\tmi-j, \tmi-\ell +j}^{Uni[0,1], \Bar{F}}(1-t_m,\ldots,1-t_{\ell+1})\Big]\Big)^2.\notag
    \end{alignat*}
\end{prop} 
As previously mentioned, 
$E_{m_1}[R\gamma^*]$ is desired to be as high as possible. As the real value of $m_1$ is not known, we require that the expected value is as high as possible over all possible values of $\tmi$, considering
\begin{equation}
\label{eq:sum_exp}
    \sum\limits_{\tmi=0}^m E_{\tmi}[R\gamma^*].
\end{equation}

For any pre-fixed family of critical vectors, we suggest computing 
this sum for different choices of the parameters and selecting the one that maximizes it.
For the considered families of critical vectors and normally distributed data, simulations indicate that Eq.~\eqref{eq:sum_exp} is maximized when using the critical vector with the largest $\lambda$ for which $\gamma^*=1$. This holds true for families of critical vectors with two parameters as well. Furthermore, the simulations indicate that the choice of maximizing Eq.~\eqref{eq:sum_exp} is reasonable, as the optimal critical vector determined in this way maximizes also $E_{m_1}[R\gamma^*]$ in most settings. These simulation results are displayed in Appendix \ref{sec:gamma_star}.

If different parameter combinations for families with two parameters have similar power, we can use the one having the smallest variance. Analogously to the case of the expected value, we consider the sum
\begin{equation}
    \label{eq:sum_var}
    \sum\limits_{\tmi=0}^m Var_{\tmi}[R\gamma^*].
\end{equation}
Then the procedure to obtain critical vectors from a family with two parameters is the following. For varying values of $\beta$, find, using a grid search algorithm, the respective value of $\lambda$ such that Eq.~\eqref{eq:sum_exp} is maximized. Then select the parameter combinations for which Eq.~\eqref{eq:sum_exp} is the maximum or reasonably close to it. Finally, among these combinations, select the one that minimizes Eq.~\eqref{eq:sum_var}.

To summarize, when choosing the critical vector, achieving high power is the main priority. For families of critical vectors with more than one parameter, different combinations may lead to the same, or similar, power. In this case, the parameter combination minimizing the variance should be chosen. 
As we will illustrate in the following section, in practical applications it can be sensible to sacrifice some power to reduce the variance, as it might increase the robustness of our methodology to violations of the model assumptions.

\section{Simulations}
\label{sec:Simulation}
In this section we discuss simulation results.
We have compared the performance of our procedure with the procedure by \cite{GoemanSolari}. We have focused on this comparison, as other approaches mentioned in Section \ref{sec:Background} are either derived from it or based on permutations and therefore defined in a different framework. First, we have studied the methods under Assumption \ref{A:model}, i.e., independence of the p-values and the correct specification of the CDF $F$. Subsequently, we have investigated the robustness of our procedure to violations of these assumptions. Finally we have explored different approaches to estimate the parameters of $F$.

We have simulated $N$ stochastically independent and identically distributed (i.i.d.) observables $\boldsymbol{X_1},\ldots,\boldsymbol{X_N}$ from a $m$-variate equicorrelated multivariate normal distribution with all marginal variances equal to one and correlation coefficient $\rho$, such that the stochastic representation $\boldsymbol{X_1}=\boldsymbol{\eta}+\boldsymbol{\epsilon}\in\mathbb{R}^m$ with $\boldsymbol{\epsilon}\sim\text{MVN}_m(\mathbf{0}, \Sigma_{\rho})$ holds true, where $\text{MVN}_m(\mathbf{0}, \Sigma_{\rho})$ denotes the $m$-variate normal distribution described before.
The entries of the vector $\boldsymbol{\eta}$ are equal to $\eta_{alt}$ for the first $m_1$ entries and zero for the remaining ones. 
We have varied $\eta_{alt}$ to account for different effect sizes $\theta$, where $\eta_{alt}=\theta / \sqrt{2}$. While the assumption of (approximately) normally distributed data is often justified in practice, often only (some) information about the effect size is available. With this, we mean that the practitioner typically neither knows the marginal variances of $\boldsymbol{X_1}$ nor that these marginal variances are all equal to each other. This necessitates Studentizing all $m$ coordinates of $X_1, \ldots, X_N$ separately in a practical data analysis situation, which we emulate here.
Furthermore, we underline that the normal distribution model is the standard model used in fMRI analysis to detect activation \citep{Lindquist2008}. For each variable we have tested the null hypothesis that the mean is zero against a two-sided alternative and we have obtained a t-statistic and the corresponding p-value via a two-sided one-sample t-test. If a null hypothesis $H_i$ is false, the t-statistic is $\tval_i\sim F_{\nu,\mu}$, where $F_{\nu,\mu}$ denotes the CDF of the t-distribution with non-centrality parameter $\mu=\theta\cdot\sqrt{N/2}=\eta_{alt}\cdot \sqrt{N}$ and $\nu=N-1$ degrees of freedom. The corresponding p-value is $P_i\sim F$, where

\begin{equation}
\label{eq:F_t_test}
    F(p)\coloneqq 1-F_{\nu,0}\left(F_{\nu,\mu}^{-1}\left(1-\frac{p}{2}\right)\right)+F_{\nu,0}\left(-F_{\nu,\mu}^{-1}\left(1-\frac{p}{2}\right)\right).
\end{equation}

Subsequently, we have applied the procedure by \cite{GoemanSolari} (``$GS$'') and the proposed method to compute a lower ($1-\alpha$)-confidence bound $\hmi$ for the number of true discoveries. For the latter, we have considered different critical vectors, chosen from the families introduced in Section \ref{sec:critical_vector}: $BH$ (Eq.~\eqref{eq:BH}), $BY$ (Eq.~\eqref{eq:BY}), $AORC$ (Eq.~\eqref{eq:AORC1}) and $Exp$ (Eq.~\eqref{eq:Exp2}). Throughout the following sections, we use the same notation to identify a family and results obtained using the family within our procedure, leaving distinction to context.

For each family, parameters were selected so that the critical vector maximizes Eq.~\eqref{eq:sum_exp} (denoted by suffix ``opt''), which in this setting always corresponds to the largest value of the parameter $\lambda$ such that $\gamma^*=1$.
To further investigate the behavior of the method, when possible we have studied additional critical vectors, taking the largest $\lambda$ such that $\gamma^*=0.95$ (denoted by suffix ``0.95''), $\gamma^*=0.9$ (``0.9'') and $\gamma^*=0.8$ (``0.8''). Families with two parameters for which different parameter combinations could lead to the same value $\gamma^*$ have been managed as suggested in the previous section. 
The performance of our methodology with different critical vectors as well as the performance of the $GS$ procedure have been studied. First, we have examined the validity of the confidence bounds, that is, checking that the proportion of iterations for which $m_1<\hat{m}_1$ is at most $\alpha$. Then we have compared results in terms of power and variability, i.e., the average and the empirical variance of the computed confidence bounds over all iterations.

All results are based on $N=50$ observations, $m=100$ hypotheses and significance level $\alpha=0.2$. We have set the number of true discoveries as $m_1\in\{5,10,20,30,$ $\ldots,90\}$. 
First, we have simulated the data and applied the procedure under the assumptions of the model, fixing the correlation parameter $\rho=0$ and assuming that the true effect size $\theta$, and thus $F$, are known. We have considered $\theta\in\{0.4, 0.6, 0.8, 1,$ $ 1.2,2,5\}$, as well as $\theta=-1$ to check whether negative effect sizes lead to different results.
Subsequently, we have analyzed the robustness of the method to violations of the independence assumption or misspecification of $F$. For the first case, we have varied the correlation parameter $\rho$. Note that $\boldsymbol{X_1},\ldots,\boldsymbol{X_N}$ remain i.i.d. under any value of $\rho$. For the second case, we have used a value $\hat{\theta}$ instead of the true effect size $\theta$ in the procedure. In practice, $\hat{\theta}$ can be an estimate determined using prior information or by estimating the effect size from external data from the same experiment or a similar one. Each scenario has been simulated $B=10,000$ times.

All simulations were carried out in \texttt{R} \citep{RCore}. We have used the \texttt{hommel} package \citep{RHommel} for $GS$, as well as the \texttt{OrdStat} package, available at \url{github.com/jvschroeder/OrdStat}, to compute the joint distribution of order statistics as in \citet{ROrdStat}. The code for the proposed method is available upon request to the authors.

The following sections show results for each scenario, while additional results and figures are reported in Appendix \ref{sec:gamma_star}. 
Since using the negative effect size $\theta=-1$ leads to similar results as the corresponding positive effect size $\theta=1$, results for $\theta=-1$ are not stated explicitly. Furthermore, $BH$ and $BY$ generally lead to similar results, therefore only results based on $BH$ are reported. Finally, for the smallest effect size ($\theta=0.4$), it was not possible to find critical vectors such that $0<\gamma^*<1$ for families with one parameter, so only results for $\gamma^*=1$ have been considered.

\subsection{Independence} \label{subsec:ind}
First we show results under the assumptions of the model, i.e., for independent p-values ($\rho=0$) and correct specification of the distribution $F$. These are discussed in terms of the resulting lower confidence bound $\hmi$: validity, power and variability.

As expected, all methods determine valid confidence bounds. Power tends to increase with the value of $\gamma^*$ used to choose the critical vector, confirming what has been stated in Section \ref{sec:critical_vector}. Hence, for each family the highest power is achieved choosing the critical vector so that $\gamma^*=1$. Moreover, for all procedures power increases with the effect size $\theta$.

\begin{figure}[ht]
    \centering
    \includegraphics[width=1\textwidth]{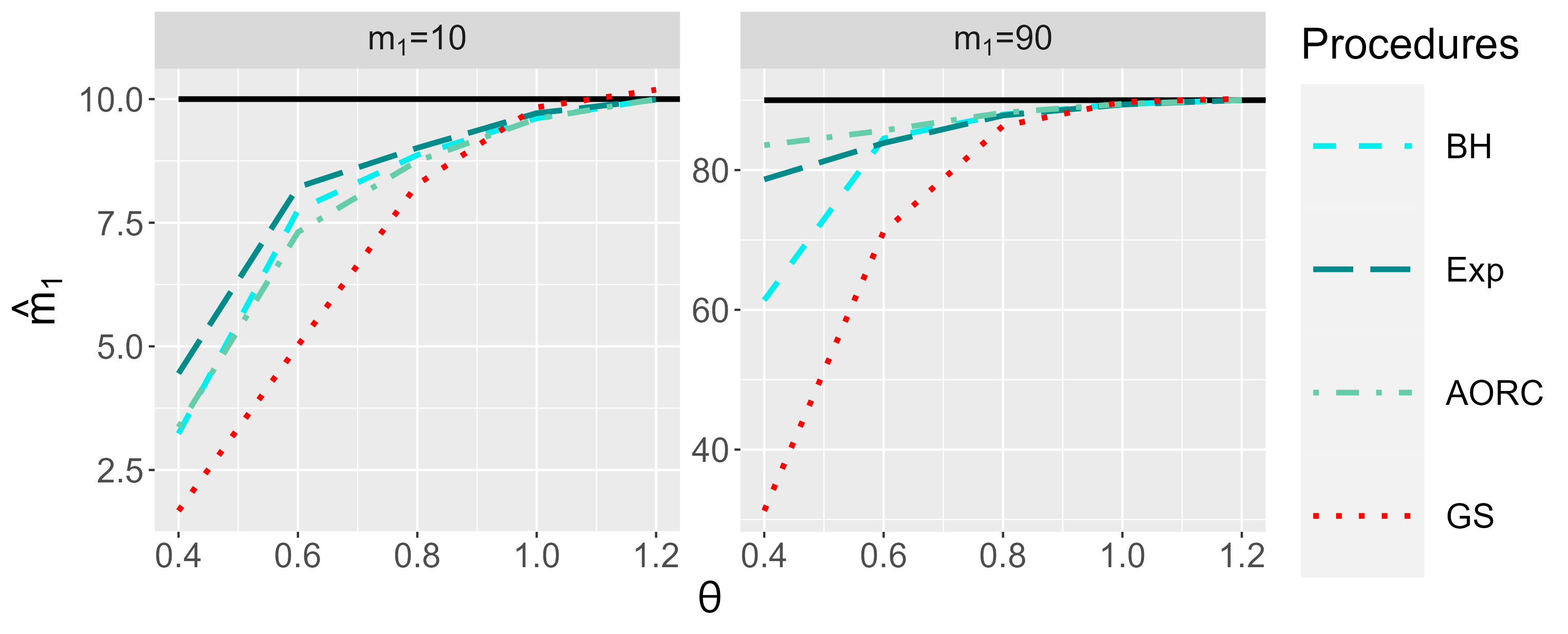}
   \caption{Independence setting: p-values have correlation $\rho=0$ and the effect size $\theta$ is known. Lower confidence bound $\hat{m}_1$ for the number of true discoveries $m_1$ obtained from the proposed procedure with different families of critical vectors ($\BH$, $\Exp$, $\AORC$) and the method of \protect{\citet{GoemanSolari}} ($GS$). The black solid line corresponds to $m_1$.}
   \label{figure:Comparison_theta_rho_0}
\end{figure}

Figure \ref{figure:Comparison_theta_rho_0} displays the average confidence bounds $\hmi$ obtained from $GS$ and the proposed method with different families and using $\gamma^*=1$. Results are shown for the smaller effect sizes ($\theta\leq 1.2$) and the extreme values of the denseness of signal ($m_1\in\{10,90\}$). For small effect sizes ($\theta\leq 1$), the proposed method using any family and $\gamma^*=1$ is more powerful than $GS$. In particular, the highest power is achieved with $\Exp$ when the signal is sparse (low $m_1$), and with $\AORC$ when it is dense (large $m_1$). $\BH$ is generally the second most powerful procedure. On the contrary, for large effect sizes $GS$ is generally the most powerful procedure, followed by $\BH$ and $\AORC$.

For all methods, the empirical variance increases as $\theta$ decreases, with a slower increase for $GS$. For small $\theta$, $GS$ has the lowest empirical variance; for the proposed procedure, the empirical variance decreases with $\gamma^*$. The opposite is true for large $\theta$: the empirical variance is higher for $GS$, and in the proposed method decreases as $\gamma^*$ increases. This indicates that there is not a linear relationship between the empirical variance and $\gamma^*$.

\subsection{Dependency among the p-values}
To investigate the robustness of our procedure to the violation of the independence assumption we have set the level of correlation $\rho\in\{0.3,0.6,0.9\}$, while $F$ is still correctly specified. Note that we have only used positive correlation coefficients because in applications such as fMRI data positive dependency of the voxels is assumed \citep{Lindquist2008}.

For larger effect sizes ($\theta\geq 1$), all methods determine valid confidence bounds for any level of correlation, except $Exp$ with $\gamma^*<1$ for large effect size ($\theta\geq 2$). As the effect size decreases, the proposed method determines valid confidence bounds using any family with decreasing $\gamma^*$.
Regarding power, for larger effect sizes ($\theta\geq 2$) it increases with $\rho$; in this setting, $GS$ remains the most powerful method. For smaller effect sizes ($\theta\leq 0.8$), the proposed method using any family with the largest $\gamma^*$ that leads to valid confidence bounds is more powerful than $GS$.
The empirical variance of all methods increases with the level of correlation $\rho$. For larger effect sizes ($\theta\geq 1.2$) this increase is faster for $GS$. Otherwise, results concerning the empirical variance are similar to the independent case ($\rho=0$). 

\subsection{Misspecification of $F$}
Since $F$ is unknown in practice and has to be determined, we now present results on the robustness of our procedure to misspecifications of $F$. When the p-values are computed from a t-test, the only unknown parameter of $F$ in Eq.~\eqref{eq:F_t_test} is the non-centrality parameter $\mu_t$, which depends on the effect size $\theta$. We have considered low values of the effect size $\theta$, for which the proposed method tends to be more powerful than $GS$ when $F$ is correctly specified (see the previous paragraphs). Then we have used different values of the assumed effect size $\hat{\theta}$, so that the true effect sizes is either overestimated ($\theta<\hat{\theta}$) or underestimated ($\theta>\hat{\theta}$). In particular, we have set $\hat{\theta}\in\{\theta-0.1, \theta-0.2\}$ for $\theta=0.8$ and $\theta=1.2$; additionally, we have used $\hat{\theta}\in\{\theta+0.1, \theta+0.2\}$ with $\theta\in\{0.4, 0.6, 0.8, 1\}$. Results are given under independence of the p-values ($\rho=0$). Note that the performance of the $GS$ procedure is not influenced by misspecification of $F$.

When $\theta$ is overestimated, all methods determine valid confidence bounds. Otherwise, confidence bounds determined using any family with $\gamma^*=1$ are no longer valid.
If $\theta$ is overestimated, the power of the proposed procedure decreases as the difference between $\theta$ and $\hat{\theta}$ increases. For smaller effect sizes ($\theta\leq 0.8$) the proposed procedure with any family and $\gamma^*=1$ is more powerful than $GS$. If $\theta$ is underestimated, the proposed method is more powerful than $GS$, using any family and the largest $\gamma<1$ for which the method determines valid confidence bounds.
Finally, the empirical variance of the proposed method with any critical vector returning valid confidence bounds increases compared to the case of correct specification of $F$.

\subsection{Dependency among the p-values and misspecification of $F$}
In practice, both assumptions of the p-value model can be violated at the same time. In this subsection we therefore show results for dependency of the p-values and misspecification of $F$. The level of correlation $\rho$ as well as the assumed and true effect sizes, $\hat{\theta}$ and $\theta$, have been fixed as in the previous subsections.
When $\theta$ is overestimated, all methods determine valid confidence bounds if the effect size is large enough ($\theta\geq 0.8$) or strongly overestimated ($\hat{\theta}\geq\theta+0.2$). 
When $\theta$ is underestimated, our procedure with any family and low $\gamma^*$ determines valid confidence bounds for any level of correlation ($\rho>0$).  
As in the previous paragraphs, we have considered only critical vectors leading to valid confidence bounds. If $\theta$ is overestimated, the proposed method with any family is more powerful than $GS$ for small values of $\theta$ and all correlations $\rho$. If $\theta$ is underestimated, on the contrary, $GS$ is generally more powerful.
When the signal is sparse (small $m_1$) and $\hat{\theta}\leq 0.8$, the empirical variance of $GS$ always increases faster than that of the proposed method as $\rho$ increases. This does not necessarily hold true when the signal is dense.

\vspace{3mm}
In summary, these simulations emphasize the importance of the choice of the critical vector for the power and robustness of our procedure. 
Our results imply that the critical vector should not always be chosen to optimize power under the independence assumption. 
As a rule of thumb, under the normal distribution model, if the effect size is large ($\theta \leq 1$) using $\gamma^*=1$ 
should lead to valid confidence bounds, otherwise it is advised to use $\gamma^*< 1$. If the effect size is so small that defining critical vectors such that $\gamma^*<1$ is impossible, it is sensible to use a constant larger than the predetermined effect size $\hat{\theta}$ to ensure the validity of the bounds. These considerations emphasize the need to find reliable estimates for the effect size.

\subsection{Estimating the effect size}
\label{sec:Simulation_effect_sizes}
The parameters of the CDF $F$ of the p-values under the alternative are generally not known.
Then, one possibility is to estimate those  parameter values on the basis of similar data. If such additional data are not available, the data set under study can be split into two parts; one part can be used to estimate the parameters, and the other to apply the method. This strategy is similar to \citet{Blain2022}, who use an additional data set to construct a family of critical vectors.

In this section we present an approach to estimate $F$ as in Eq.~\eqref{eq:F_t_test} when each p-value $P_i$ is derived from a one-sample two-sided t-test, and thus from a t-statistic $\tval_i$, as in the setting of our simulations. All the t-test statistics and p-values, respectively, are assumed to be independent.
The only unknown parameter in the model described at the beginning of this section is the non-centrality parameter $\mu$, as the degrees of freedom $\nu=N-1$ only depend on the sample size. 
All other relevant quantities, including the CDF $F$ of the p-values under the alternative, are determined by $\mu$ and $\nu$ by virtue of Eq.~\eqref{eq:F_t_test}.

If we knew which t-statistics correspond to false null hypotheses (e.g., $\tval_1,\ldots,\tval_{m_1}$), we could use them to easily compute an estimate $\hat{\mu}$ of $\mu$. Indeed,
\begin{equation*}
    E[\tval_i]=\mu\cdot\sqrt{\frac{\nu}{2}}\left(\frac{\Gamma((\nu-1)/2)}{\Gamma(\nu/2)}\right)\qquad (i=1,\ldots,m_1)
\end{equation*}
and so
\begin{equation}
\label{eq:est_ef_size}
    \hat{\mu}= \Bar{\tobs}\cdot\sqrt{\frac{2}{\nu}} \left(\frac{\Gamma(\nu/2)}{\Gamma((\nu-1)/2)}\right),\qquad \Bar{\tobs}=\frac{1}{m_1}\sum_{i=1}^{m_1} \tobs_i
\end{equation}
where $\tobs_i$ is the observed value of $\tval_i$. Then the effect size $\theta$ could be immediately estimated with $\hat{\theta}=\hat{\mu}\cdot\sqrt{2/N}$.

In practice, it is unknown which null hypotheses are false.Therefore, we suggest to compute t-statistics for all $m$ coordinates and use a thresholding scheme to decide which coordinates are considered as alternative.
A first intuitive strategy is selecting those t-statistics that are not smaller than a given threshold, which may be chosen, for instance, as the empirical $\omega$-quantile of all the t-statistics.
A second, similar strategy is selecting the t-statistics for which the corresponding p-value does not exceed a threshold.
This threshold may be either a fixed value $c$ or the threshold of a single-step test controlling the FWER, i.e., the probability of selecting at least one t-statistic corresponding to a true null hypothesis. We consider the Bonferroni and \v{S}id\'{a}k corrections controlling the FWER at level $a$ \citep[see][Section 5]{dickhaus_2014}. However, other thresholds could be utilized as well. Notice that the threshold increases with $a$. An algorithm for this second strategy is given in Appendix \ref{sec:Algorithm}.

We have used the previous simulation setting to study the performance of different selection criteria. Based on the results from the previous simulation study, the goal is finding precise estimators while avoiding underestimation of the effect size. We have considered the thresholds for the t-values based on the $\omega$-quantiles with $\omega\in\{25\%,50\%\}$; then the thresholds for the p-values based on fixed $c\in\{0.01,0.1\}$ and on FWER control with level $a\in\{0.01,0.05,0.1\}$.
When no t-values were selected, the effect size has been set to zero. In practice, this would lead to a valid TDP confidence bound equal to zero, as all null hypotheses would be assumed to be true.

Generally, the estimated effect size increases with the denseness of the signal while the empirical variance of all estimators decreases. Using thresholds for the t-values leads to general underestimation of the effect size. Therefore this strategy is not suitable.

The performance of thresholds for the p-values depends on the true effect size $\theta$. For small effect sizes, large  threshold values rarely lead to an underestimate. Using small values often leads to an estimated effect size of zero; otherwise, the effect size is generally overestimated.
As the true effect size increases, both strategies may underestimate the effect size. This is more severe for large values of the thresholds (i.e., large values of $c$ and $a$) and in general more frequent when using the fixed thresholds $c$. 
When the level of correlation $\rho$ increases, fixed thresholds $c$ lead to less underestimation than FWER-based thresholds.

These simulations support the use of thresholds for the p-values. The value of the threshold should be large when the true effect size $\theta$ is small, and small when $\theta$ is large. In practice we would therefore recommend to estimate the effect size using FWER-based thresholds with small values of $a$ (e.g. $a\leq0.01$) if the data are assumed to be normally distributed and a two-sided one-sample t-test is applied. If this leads to an effect size of zero, the threshold can be increased.

\section{fMRI study}
\label{sec:Case_Study}
In this section, we present results obtained from the analysis of a real fMRI data set. The aim of task-related fMRI is to identify regions in the brain that react to a stimulus. To do so, researchers measure the blood oxygenation level dependent (BOLD) signal in different voxels and at different time points; the experiment is usually repeated for several subjects. Multi-subject data in the context of fMRI are typically analyzed using two-level mixed-effects models \citep{Lindquist2008}. At the first level, data of the individual subjects are analyzed. The output of the first level is used as input of the second level, which is the group analysis. Subjects are then considered as random effects \citep{Poldrack2011}.
Suppose that the study considers $m$ voxels, $W$ time points and $N$ subjects. Furthermore, suppose that the stimuli of interest can be coded using $L$ conditions.
In a first level analysis, the pre-processed BOLD signal is fitted to a theoretical BOLD signal using a generalized linear model (GLM) :
\begin{equation*}
    \Tilde{\boldsymbol{Y}}_{ij}=\boldsymbol{X}_{ij}\cdot\boldsymbol{\beta}_{ij}+\boldsymbol{\epsilon}_{ij}\qquad (i=1,\ldots,m;\,j=1,\ldots, N)
\end{equation*}
where $\Tilde{\boldsymbol{Y}}_{ij} \in\mathbb{R}^W$ represents the pre-processed time series of the observed BOLD response for the $i$th voxel and $j$th subject. The design matrix $\boldsymbol{X}_{ij}\in\mathbb{R}^{W\times K}$ accounts for the theoretical BOLD signal of the $L$ conditions as well as nuisance parameters, such that $K\geq L$. Finally, $\boldsymbol{\beta}_{ij}=(\beta_{ij1},\ldots,\beta_{ijK})^\top$ is the vector of coefficients and $\boldsymbol{\epsilon}_{i,j}\in\mathbb{R}^W$ is assumed to follow an AR(2)-process \citep{Lindquist2008}.

The beta coefficients are used in the second level analysis to find voxels reacting to the stimulus of interest. They are considered to be random variables. Often, researchers want to find differences in brain activation between two stimuli, a so-called contrast. In this case, for each subject the difference between two stimulus-specific beta coefficients is computed, that is
\begin{equation*}
    D_{ij}=\beta_{ij1}-\beta_{ij2},\qquad D_{ij}=\mu_i +\xi_{ij},
\end{equation*} where $D_{ij}$ denotes the contrast of the $i$th voxel and $j$th subject and $\xi_{ij}$ is normally distributed with mean zero.
These values are then used in the group analysis to test at each voxel $i$ the null hypothesis $H_0:\mu_i=0$, using a one-sample two-sided t-test \citep{Lindquist2008}. Therefore, the entire analysis produces a map of t-statistics $\tval_1,\ldots,\tval_m$ and the corresponding map of p-values $P_1,\ldots,P_m$.
There are over $100{,}000$ voxels within the brain \citep{Lindquist2008} and inference on sets of contiguous voxels is of interest. Given a pre-defined region of interest (ROI), that is, a pre-defined set of contiguous voxels, and the p-values for the corresponding voxels, our procedure computes a lower confidence bound for the proportion of active voxels (TDP). The pipeline for the entire analysis is illustrated in Appendix \ref{sec:Analysis_fMRI}.

We have considered fMRI studies which investigate the difference between listening to sounds produced by human voice versus non-human sounds. ROIs have been defined based on \citet{Schirmer2012} and \citet{Binder2000}. 
The shape and location of the ROIs have been chosen based on the reported peaks of activation and size of the clusters. In total, we have analyzed six ROIs, two located in the left hemisphere, the left Superior Temporal Gyrus (L STG) and left Auditory Cortex (L AC), and four located in the right hemisphere, the right Auditory Cortex (R AC), the right  Fusiform Gyrus (R FG), the right Superior Temporal Gyrus (R STG) and right Middle Temporal Gyrus (R MTG). Further information on the definition and attributes of the ROIs are given in Appendix \ref{sec:Analysis_fMRI}.

We have studied fMRI data collected by \citet{Pernet2015} and available at 
\url{https://openneuro.org/datasets/ds000158/versions/1.0.0} to compute \\
confidence bounds for the TDP. In this experiment, participants passively listened to a human voice versus non-human sounds. 
Pre-processed data and contrast maps as results of the first-level analysis for 140 subjects are available at \url{https://github.com/angeella/fMRIdata/tree/master/data-raw/AuditoryData} and have \\ been used in this study. Further information about pre-processing and first level analysis can be found in \citet{Andreella2023}. For each pre-specified ROI, we have applied our methodology as follows. Since the procedure requires knowledge about (the parameters of) $F$, we have randomly split the subjects in two subgroups to  estimate the effect size ($N^{[e]}=50$) and compute the confidence bounds ($N^{[b]}=90$). Within each of the two subsets, we have tested the null hypothesis for each voxel in the ROI that there is no difference between the activation caused by human voice vs.~non-human sounds and obtained t-value and p-value maps, using the \texttt{pARI} package in \texttt{R} \citep{pARI_manual}.

The effect sizes for the ROIs have been estimated using Eq.~\eqref{eq:est_ef_size}. As suggested in the simulation study, we have used a p-value threshold based on the \v{S}id\'{a}k correction with $a=0.01$. Since this led to an estimated effect size of zero for the ROI ``R FG'', we have used the fixed threshold $c=0.1$ for this ROI. The estimated effect size within each ROI are displayed in table \ref{table:Result_auditory}.

To compute $80\%$-confidence bounds for the TDP, we have considered the families introduced in Section \ref{sec:critical_vector} and, for each ROI and each family, we have selected a critical vector based on the effect size and suggestions given in Section \ref{sec:Simulation}. For ROIs with large effect sizes ($\hat{\theta}\geq 1.2$), we have fixed the critical vectors to optimize Eq.~\eqref{eq:sum_exp}, that is $\gamma^*=1$. 
For ``R AC'' we have taken $\gamma^*=0.95$. Finally, since ``R FG'' has a very small effect size ($|\hat{\theta}|<0.4$) we have replaced the effect size by the constant 0.5 and considered $\gamma^*=1$.
Based on the effect size and critical vectors, we have then computed the lower confidence bounds for the TDP. 

Table \ref{table:Result_auditory} contains the confidence bounds obtained using the proposed procedure with different families, as well as the confidence bounds based on the method by \citet{GoemanSolari}. Within each ROI, different choices of the family lead to similar result. Furthermore, these are equivalent to $GS$ for ROIs with large (estimated) effect size ($\hat{\theta}>1$). For the cluster ``R AC'', the proposed procedure based on the $Exp$ family with $\gamma^*=0.95$ determines the largest bounds. The proposed method based on the $BH$, $BY$ and $AORC$ family with $\gamma^*=0.95$ determine confidence bounds smaller than the $GS$ procedure. For the cluster ``R FG'', our procedure returns higher bounds than $GS$. This is in accordance with the results of the simulation study, which indicated that our procedure has similar power to $GS$ for larger effect sizes and is more powerful than $GS$ for smaller effect sizes. 

These result suggest that most of the ROIs found by \citet{Schirmer2012} and \citet{Binder2000} are well defined and truly relevant for human voice processing in contrast to non-verbal sound processing. Even though we have explored different ROIs, it has to be noted that the proposed procedure is constructed to study a single, pre-fixed region of interest. Furthermore, the validity of the resulting confidence bounds is based on the assumption that the effect size is not underestimated and that the choice of $\gamma^*$ accounts for possible correlations between voxels. This emphasizes the importance of choosing appropriate values for both $\hat{\theta}$ and $\gamma^*$.

\section{Discussion}
We have presented a procedure to compute lower confidence bounds for the TDP for a fixed set of null hypotheses. In contrast to established procedures, our method utilizes a step-up multiple testing procedure and computes the confidence bounds based on the distribution of the respective number of rejections. The proposed method assumes that the p-values corresponding to the null hypotheses of interest are stochastically independent and that the distribution of the p-values under the alternative is known. We have demonstrated through simulations that the proposed procedure is more powerful than the established method of \citet{GoemanSolari} in certain scenarios and, if the settings of the procedure are chosen properly, it can be robust against some violations of the assumptions. Then we have demonstrated how to use the method for fMRI data analysis, when inference is made on a region of interest defined a priori, e.g., to test the reproducibility of a study or to explore if a region associated with a certain task is active when performing another task.

The proposed method is based on a generic choice of the critical vector of a step-up procedure. The observed number of rejections as well as a quantity $\gamma^*$, which depends on the distribution of the random number of rejections, are used to compute the confidence bounds. While the procedure defines a valid confidence bound for any choice of the critical vector, the choice influences the power of the proposed procedure.
We have given guidelines on how to select an appropriate critical vector from a pre-fixed family indexed by a parameter $\lambda$ (and in some cases additional parameters). In particular for normally distributed data we have illustrated that in most cases power is maximized for the critical vector having the largest $\lambda$ such that $\gamma^*=1$.

If the data are normally distributed and the aforementioned assumptions are met, we have demonstrated through simulation that, for small effect sizes, the proposed procedure can be more powerful than the method of \citet{GoemanSolari}. For larger effect sizes, the method by \citet{GoemanSolari} is slightly more powerful but results have higher variability. Furthermore, we have investigated the robustness of the proposed procedure against violations of the assumptions. For large effect sizes, the proposed method is robust to positive dependency among the p-values. If the effect size is small, critical vectors should be chosen such that $\gamma^*<1$ to obtain valid confidence bounds. 
The computation of the confidence bounds utilizes information about the effect size, which can be determined using either prior information or estimation based on external information. If the true effect size is overestimated in absolute value, i.e., the true effect size is smaller than the assumed effect size in absolute value, the proposed procedure using any family and $\gamma^*\leq 1$ returns valid confidence bounds. This cannot be ensured if the true effect size is underestimated, again in absolute value. This emphasizes the need for a good estimation procedure for the effect size.

Future work will focus on extending the proposed procedure to account for dependency among the p-values. To do so, the distribution of the number of rejections under dependency of the p-values is needed. Some approaches are given in \citet{RoquainVillers} and would have to be adapted to fit our requirements. Another possible direction for future research consists in obtaining simultaneity of the proposed methods over several (or even all possible) subsets of an original set of null hypotheses. In the fMRI example, this original set of null hypotheses could for instance refer to all measured voxels. Conceptually, this simultaneity can be achieved by applying the closed testing principle and carrying out our proposed method for every intersection null hypothesis in the closure (with respect to intersections) of the original set of null hypotheses. However, proceeding (naively) in this manner will induce a huge computational effort, such that computational shortcuts are desirable. The development of such shortcuts appears challenging and interesting at the same time. Finally, it appears desirable to analyze the choice of the tuning parameters of our procedure also from a theoretical point of view (complementing the numerical evidence from computer simulations), and for more general models than those assuming normally distributed data.

\section{Funding acknowledgement}
Friederike Preusse gratefully acknowledges funding by the Deutsche Forschungsgemeinschaft (DFG, German Reseach Foundation)- project number 281474342.\\
Anna Vesely and Thorsten Dickhaus acknowledge financial support by the Deutsche Forschungsgemeinschaft (DFG) via Grant No.~DI 1723/5-3.

\section{Data availability}
The data used in this article are available at the OpenNeuro dataset ds000158 at \url{https://openneuro.org/datasets/ds000158/versions/1.0.0} (raw data) and \url{https://github.com/angeella/fMRIdata/tree/master/data-raw/AuditoryData} (pre-processed data). The code for the simulation and analysis is available upon request from the authors.
\newpage
\clearpage
\bibliography{quellen}
\newpage
\section{Tables}
\begin{table}[ht]
\caption{fMRI data: analysis of different ROIs. Number of voxels $m$, estimated effect size $\hat{\theta}$ and lower confidence bound $\hat{\pi}_1$ for the TDP obtained from the method of \citet{GoemanSolari} ($GS$) and the proposed procedure with different families of critical vectors ($\BH[\gamma^*]$, $\BY[\gamma^*]$, $\AORC[\gamma^*]$, $\Exp[\gamma^*]$). The value $\gamma^*$ was chosen based on the suggestions given in Section \ref{sec:Simulation}.}
\label{table:Result_auditory}
\centering
\begin{tabular}{lrrllllll}
  \hline
ROI & Size &Effect size & \multicolumn{5}{l}{ $\hat{\pi}_1$ computed using: } \\ 
&|voxel|& & GS & $\BH[\gamma^*]$ & $\BY[\gamma^*]$ & $\AORC[\gamma^*]$ &  $\Exp[\gamma^*]$\\
  \hline
L STG &162& 1.3222& 1&1&1&1&1\\
L AC & 33&1.4440 & 1 &1&1&1&1\\
R AC & 257&0.9400 & 0.4825& 0.4397& 0.4397&0.4591 & 0.5175\\
R FG & 81&-0.3444 &0.1605 &0.2222&0.2222&0.2222&0.2222 \\ 
R MTG &257&1.5204 &1 &1 &1 &1 &1 \\
R STG & 389&1.6570&1 & 1& 1&1 &1  \\
   \hline
\end{tabular}
\end{table}

\section{Figures}

\newpage

\appendix

\section{Proofs} \label{sec:allproofs}

\subsection*{Theorem \ref{theorem:lower_bound}}
\label{sec:proof_lb}
The number of true discoveries $m_1$ is an unknown integer between $0$ and $m$. Fix a candidate value $\tmi\in\{0,\ldots,m\}$, as well as the probability $\probfm[\tmi]$ under the corresponding model. From Eq.~\eqref{eq:gammatmi},
the value $\gamma_{\tmi}$ is such that
\[\probfm[\tmi](\tmi\geq R\gamma_{\tmi})\geq 1-\alpha.\]
Then consider the minimum $\gamma^*$ of these values, as defined in Eq.~\eqref{eq:hmi}.
For any $\tmi$, we have that $R\gamma^*\leq R\gamma_{\tmi}$, and so
\[\probfm[\tmi](\tmi\geq R\gamma^*)\geq\probfm[\tmi](\tmi\geq R\gamma_{\tmi})\geq 1-\alpha .\] 
This holds in particular for the true value $m_1$, so that
\[\probfm(m_1\geq R\gamma^*)\geq 1-\alpha.\]
Finally, recall that $m_1$ can take only integer values. Hence
\[m_1\geq R\gamma^* \quad\Longleftrightarrow\quad m_1\in \{\lceil R\gamma^*\rceil ,\ldots, m\} \quad\Longleftrightarrow\quad m_1\geq \lceil R\gamma^*\rceil \]
and so
\[\probfm(m_1\geq \lceil R\gamma^*\rceil)=\probfm(m_1\geq R\gamma^*)\geq 1-\alpha.\]
Therefore $\hmi=\lceil r\gamma^*\rceil$ is a lower $(1-\alpha)$-confidence bound for $m_1$.

\subsection*{Theorem \ref{theorem:compute_gamma}}
\label{sec:proof}
From Eq.~\eqref{eq:gammatmi}
\begin{equation*}
    \gamma_{\tmi}=\max\left\{\gamma\in [0,1]\,:\,\probfm[\tmi](R\gamma\leq \tmi)\geq 1-\alpha\right\}.
\end{equation*}
Recall that $R$ is a discrete variable taking values in $\{0,\ldots,m\}$, so it is sufficient to study its CDF in these values.
First, suppose that $\tmi=0$. In this case,
\begin{align*}
\probfm[0](R\gamma \leq 0)=\probfm[0](R\gamma =0)=
\begin{cases}
1\quad\text{if }\gamma=0\\
\probfm[0](R=0)\quad\text{if }\gamma\in (0,1].
\end{cases}
\end{align*}
Hence it is sufficient to look for the maximum in $\{0,1\}$:
\begin{align*}
\gstar[0]=\max\left\{\gamma\in \{0,1\}\,:\,\probfm[0]\left( R\gamma = 0\right)\geq 1-\alpha\right\}=
\begin{cases}
1\quad\text{if } \probfm[0](R= 0)\geq 1-\alpha\\
0\quad\text {otherwise}
\end{cases}
\end{align*}
Then consider any other value $\tmi\in \{1,\ldots,m\}$. Notice that for $\gamma=\tmi/m$ we obtain
\begin{equation*}
    \probfm[\tmi](R\gamma\leq \tmi)=\probfm[\tmi](R\leq m)=1\geq 1-\alpha,
\end{equation*}
and so $\gamma_{\tmi}\geq\tmi/m >0$. Then we can re-write
\begin{align*}
\gstar&=\max\left\{\gamma\in \left[\frac{\tmi}{m},1\right]\,:\,\probfm[\tmi]\left(R\leq \frac{\tmi}{\gamma}\right)\geq 1-\alpha\right\}\\
&=\max\left\{\gamma\in \left\{\frac{\tmi}{m},\frac{\tmi}{m-1},\ldots,\frac{\tmi}{\tmi}\right\}\,:\,\probfm[\tmi]\left(R\leq \frac{\tmi}{\gamma}\right)\geq 1-\alpha\right\}\\
&=\max\left\{\gamma=\frac{\tmi}{\ell},\; \ell\in \{\tmi,\ldots,m-1,m\}\,:\,\probfm[\tmi](R\leq \ell)\geq 1-\alpha\right\}\\
&=\frac{\tmi}{\ell_{\tmi}}
\end{align*}
where
\begin{equation*}
    \ell_{\tmi}=\min\left\{\ell\in\{\tmi,\ldots,m-1,m\}\,:\,\probfm[\tmi](R\leq \ell)\geq 1-\alpha\right\} .
\end{equation*}
From this, it follow that
\[\gamma_{\tmi} = \tmi / \ell_{\tmi}.\]

\subsection*{Proposition \ref{prop:distribution_R}}
\label{sec:lemmas}

Denote by $S=|\mathcal{R}\cap \mathcal{M}_1|$ the random variable corresponding to the number of true discoveries in the rejection set $\mathcal{R}$, and by $V=R-S$ the random variable corresponding to the number of false discoveries. We use that $R=V+S$, then by the law of total probability
\begin{align*}
    \probfm[\tmi](R=\ell)&=\sum_{j=0}^\ell\probfm[\tmi](V=j, S=\ell-j)\\
    &= \sum_{j=0}^\ell\binom{m-\tmi}{j}\binom{\tmi}{\ell-j}t_{\ell}^j(F(t_{\ell}))^{\ell-j}\\
    &\qquad \qquad\cdot\Psi_{m-\tmi-j,\tmi-\ell+j}^{Uni[0,1], \overline{F}}(1-t_m,\dots,1-t_{\ell+1}),\notag
\end{align*} The last equality is due to \cite{RoquainVillers}, Section 5.3. The expression for $\probfm[\tmi](R=\ell)$ given in the proposition follows.

\subsection*{Proposition \ref{Prob:Var_Mean}}
\label{sec:proof_e_v}
Because $R$ is discrete, $\probfm[\tmi](R=\ell)=\probfm[\tmi](R\leq \ell)-\probfm[\tmi](R\leq \ell-1)$. Thus, the expected value of $R$ can be computed as
\begin{equation*}
    E_{\tmi}[R]=\sum\limits_{\ell=0}^m \ell\cdot \probfm[\tmi](R=\ell)=\sum\limits_{\ell=0}^m l\cdot (\probfm[\tmi](R\leq \ell)-\probfm[\tmi](R\leq \ell)),
\end{equation*}
where $\probfm[\tmi](R\leq \ell)$ is given in Proposition \ref{prop:distribution_R}.
The expression of $ E_{\tmi}[R\gamma^*]$ follows.

Similarly, the variance is computed as
\begin{align*}
    &Var_{\tmi}[R]=E_{\tmi}[R^2]-E_{\tmi}[R]^2\\
    &=\sum\limits_{\ell=0}^m \ell^2\cdot (\probfm[\tmi](R\leq \ell)-\probfm[\tmi](R\leq \ell-1))-\left(\sum\limits_{\ell=0}^m \ell\cdot (\probfm[\tmi](R\leq \ell)-\probfm[\tmi](R\leq \ell-1))\right)^2.
\end{align*}

\section{Algorithms}
\label{sec:Algorithm}
In this section, we give algorithms for the proposed methodology.

Algorithm \ref{algorithm:TDP_1} computes a lower ($1-\alpha$)-confidence bound $\hat{m}_1$ for the number of true discoveries as in Eq.~\eqref{eq:def_bound2},
using results from Theorems \ref{theorem:lower_bound}
and \ref{theorem:compute_gamma}
and Proposition \ref{prop:distribution_R}.
Note that observations in form of $r$ are only used in the last step of the algorithm.

\begin{algorithm}
\caption{Algorithm to compute $\hat{m}_1$ such that $\mathbb{P}_{m_1}(m_1\geq \hmi)\geq 1-\alpha$ as in Eq.~\eqref{eq:def_bound2}.}
\For{$\tmi=0,\dots,m$}{
    $\ell =-1$ \;
    $Pr = 0$ \;
    \While{$Pr<1-\alpha$ \& $\ell<m$}{
        $\ell=\ell+1$ \;
        $Pr=Pr+P(R=\ell)$ \;
    }
    \eIf{$\ell<\tmi$}{$\ell_{\tmi}=\tmi$\;
    }{$\ell_{\tmi}=r$}
    \eIf{$\ell_{\tmi}=0$}{$\gamma_{\tmi}=1$\;}{$\gamma_{\tmi}=\frac{\tmi}{\ell_{\tmi}}$}
    } 
$\gamma^*=\min_{\tmi\in[m]}(\gamma_{\tmi})$\;
\Return $\lceil\tr\cdot\gamma^*\rceil$\;
\label{algorithm:TDP_1}
\end{algorithm}

Algorithm \ref{algorithm:effectsize} implements the methods based on thresholds for the p-values to compute an estimate $\hat{\theta}$ of the effect size, as described in Section \ref{sec:Simulation_effect_sizes}.
The algorithm uses t-values $\mathbf{\tobs}=(\tobs_1,\ldots,\tobs_m)^\top$ and p-values $\boldsymbol{p}=(p_1,\ldots,p_m)^\top$. It selects the t-values for which the corresponding p-values do not exceed a given threshold $h$, and uses those to determine $\hat{\theta}$. The threshold $h$ may be either a fixed value or the threshold based on of a single-step test controlling the FWER at level $a$, such as the Bonferroni and \v{S}id\'{a}k corrections. For the Bonferroni correction,
\begin{equation*}
    h = \frac{a}{m},
\end{equation*}
while for the \v{S}id\'{a}k correction,
\begin{equation*}
    h = 1-(1-a)^{\frac{1}{m}}.
\end{equation*}

\begin{algorithm}
\caption{Algorithm to compute $\hat{\theta}$ based on Eq.~\eqref{eq:est_ef_size},
using a threshold $h$ for the p-values.}
$\nu=N-1$\;
$\mathbf{\tobs}_{\text{sel}}=\mathbf{\tobs}[\mathbf{p}\leq h]$\;
$\hat{\mu}=\text{mean}(\mathbf{\tobs}_{\text{sel}})\cdot\sqrt{\frac{2}{\nu}}\frac{\Gamma(\nu/2)}{\Gamma((\nu-1)/2)}$\;
\Return $\hat{\mu}\cdot\sqrt{\frac{2}{N}}$
\label{algorithm:effectsize}
\end{algorithm}

\section{Additional simulation results}
\label{sec:gamma_star}
In this section, further details about the simulations of Section \ref{sec:Simulation}
are given. We claim that, for any family of critical vectors indexed by a parameter $\lambda$ (and eventually other parameters), optimal power is achieved when selecting the largest $\lambda$ for which $\gamma^*=1$ in Eq.~\eqref{eq:hmi}
, as mentioned in Section \ref{sec:critical_vector}.
Subsequently, we present additional plots that support the claims on results made in Section \ref{sec:Simulation}.
Tables and plots are shown only for some choices of the parameters, but other values lead to the same conclusions.

First, consider the setting under Assumption \ref{A:model},
where p-values are independent and the CDF $F$ is correctly specified. Table \ref{table:theta_sum_exp} displays the sum of the expected values given in Eq.~\eqref{eq:sum_exp},
obtained using different critical vectors and for varying effect sizes $\theta$. Eq.~\eqref{eq:sum_exp}
is maximized when $\lambda$ is chosen as the largest value for which $\gamma^*=1$. Furthermore, Table \ref{table:theta_08_rho_0} displays the average confidence bounds of the proposed procedure with different critical vectors for a fixed $\theta=0.8$ and varying values of $m_1$. In most cases, the largest confidence bound is obtained for critical vectors with $\gamma^*=1$, with the only exception of settings with very sparse signal ($m_1\leq10$). This justifies the choice of maximizing Eq.~\eqref{eq:sum_exp}
to select the critical vector, as in most cases the same critical vector leads to the largest confidence bound.

\begin{figure}[p]
    \centering
    \includegraphics[width=1\textwidth]{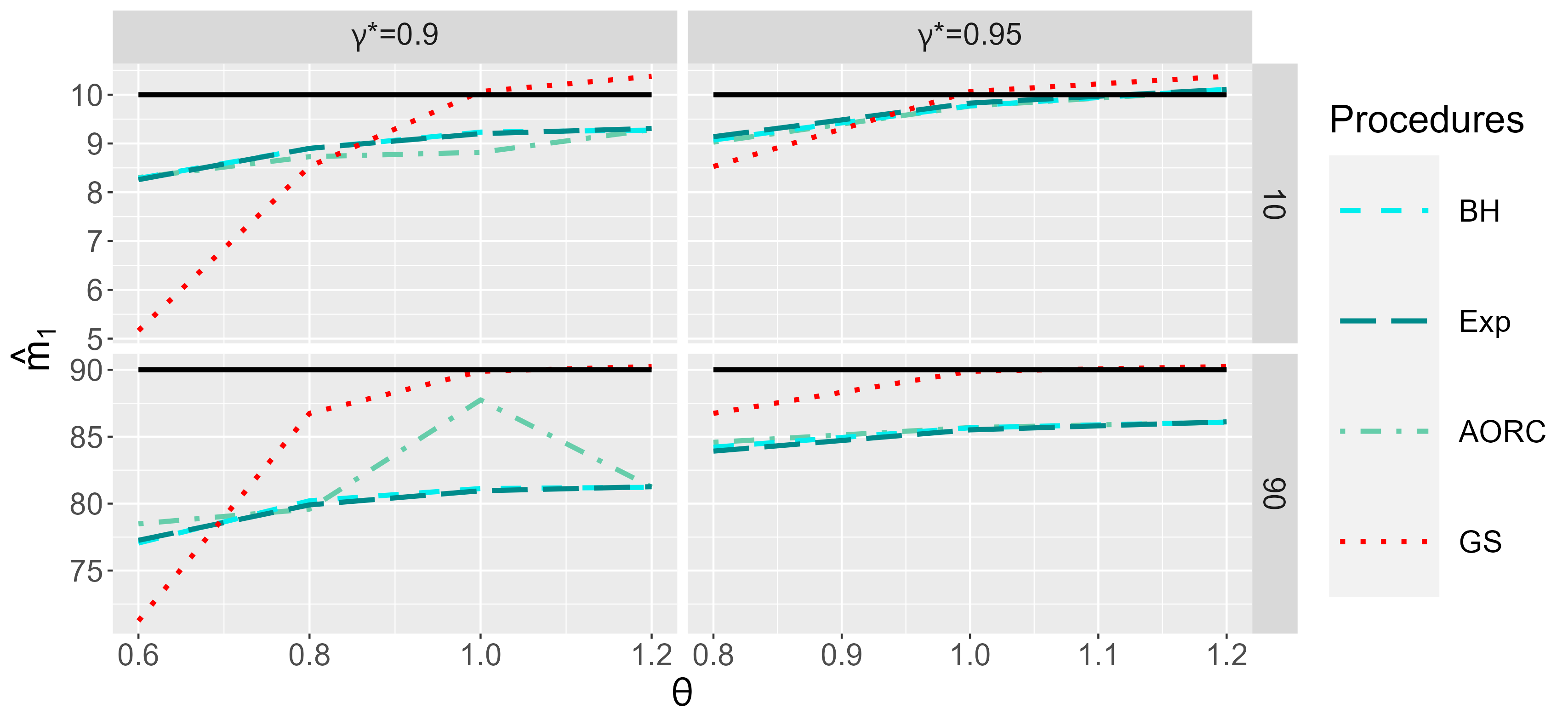}
   \caption{Dependency: p-values have correlation $\rho=0.3$ and effect size $\theta$ is known. Lower confidence bound $\hat{m_1}$ obtained from the proposed procedure with different critical vectors ($\BH[\gamma^*]$,$\Exp[\gamma^*]$, $\AORC[\gamma^*]$) and the method of \protect{\citet{GoemanSolari}} ($GS$). The black solid line corresponds to $m_1$.}
   \label{figure:Comparison_theta_rho_03}
\end{figure}

\begin{figure}[p]
    \centering
    \includegraphics[width=1\textwidth]{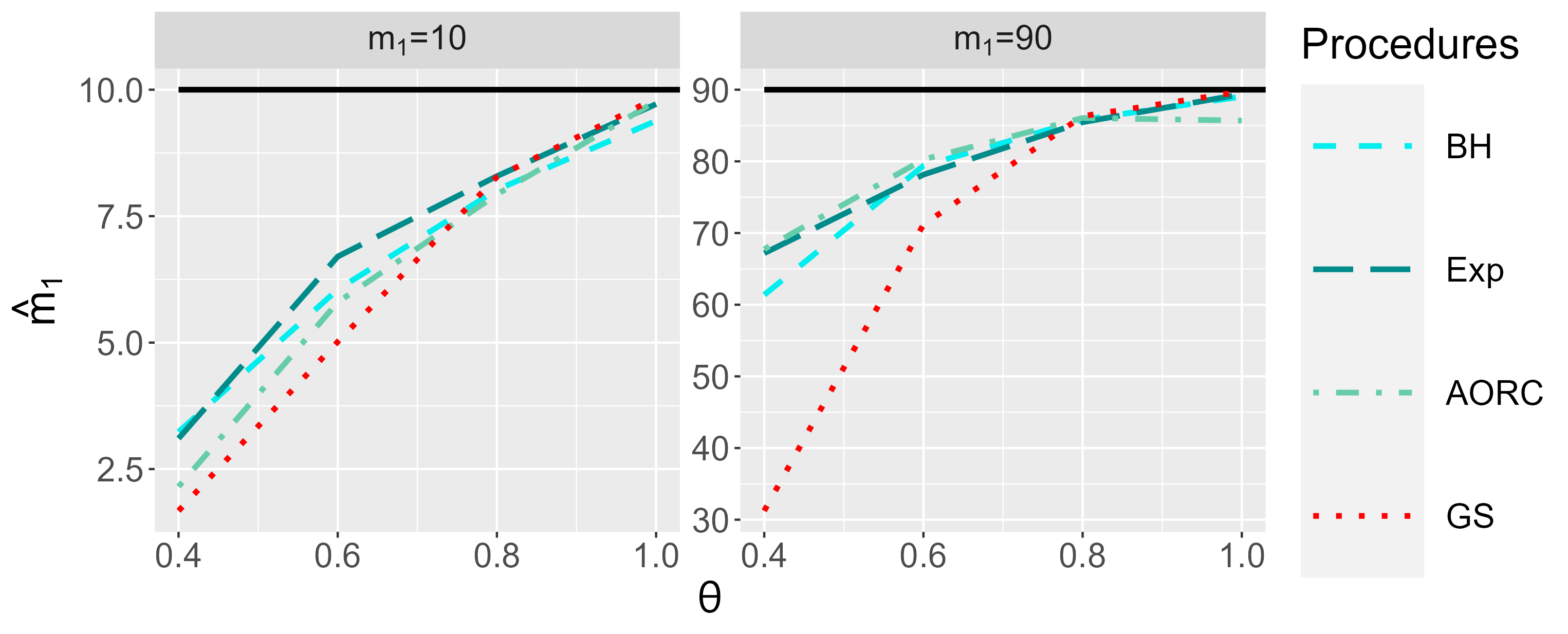}
   \caption{Misspecification of $F$: p-values have correlation $\rho=0$ and the true effect size is overestimated ($\hat{\theta}=\theta+0.1$). Lower confidence bound $\hat{m_1}$ for the number of true discoveries $m_1$ obtained from the proposed procedure with different critical vectors ($\BH$, $\Exp$, $\AORC$) and the method of \protect{\citet{GoemanSolari}} ($GS$). The black solid line corresponds to $m_1$.}
   \label{figure:Comparison_theta_mis_rho_0}
\end{figure}

\begin{figure}[p]
    \centering
    \includegraphics[width=1\textwidth]{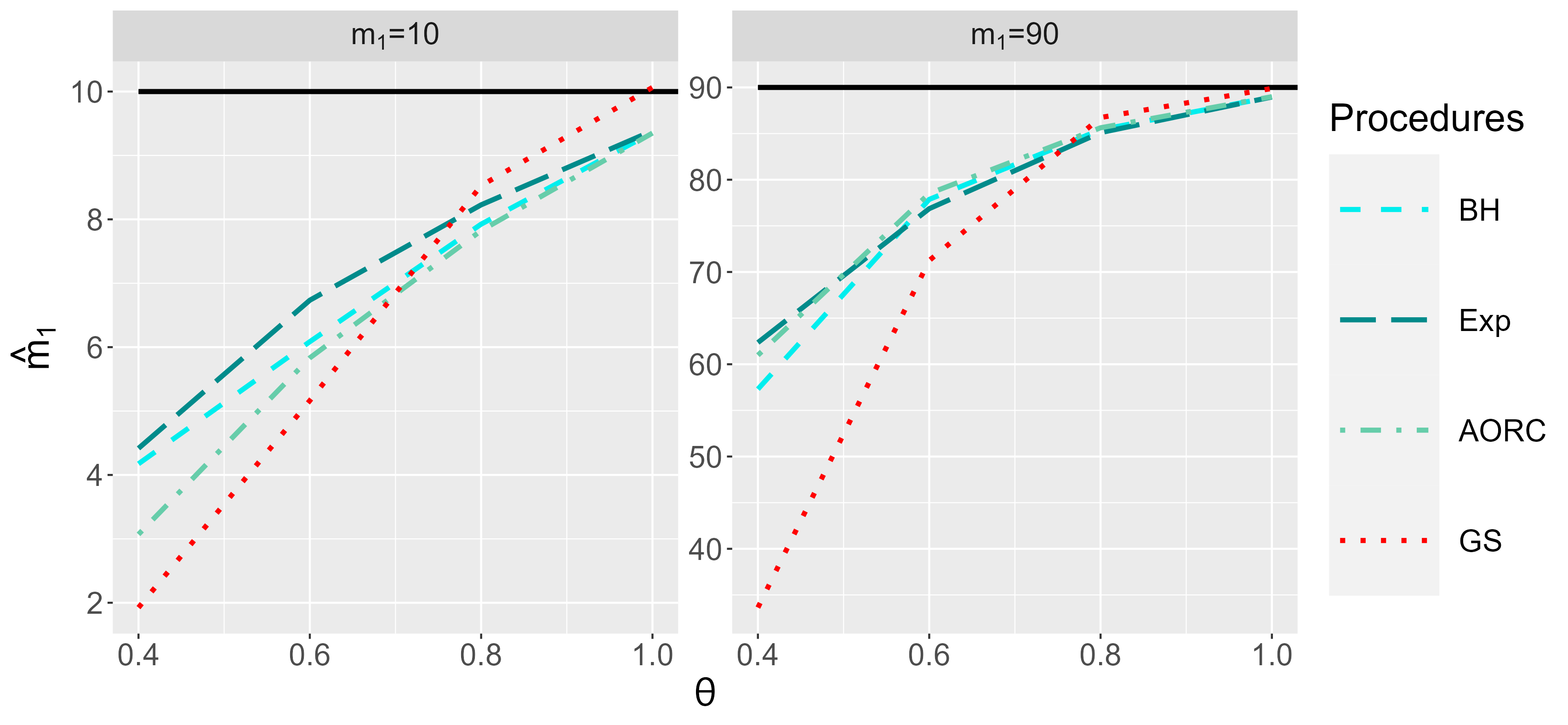}
   \caption{Dependency and misspecification of $F$: p-values have correlation $\rho=0.3$ and the true effect size is overestimated ($\hat{\theta}=\theta+0.1$). Lower confidence bound $\hat{m_1}$ for the number of true discoveries $m_1$ obtained from the proposed procedure with different critical vectors ($\BH$, $\Exp$, $\AORC$) and the method of \protect{\citet{GoemanSolari}} ($GS$). The black solid line corresponds to $m_1$.}
   \label{figure:Comparison_theta_mis_rho_03}
\end{figure}

Figures \ref{figure:Comparison_theta_rho_03}-\ref{figure:Comparison_theta_mis_rho_03} display, for different simulation scenarios, the average lower confidence bounds $\hmi$ obtained from $GS$ and the proposed method with different families of critical vectors.
Critical vectors of each family have been chosen using the largest $\gamma^*\in\{0.8,0.9,0.95,1\}$ such that 
the proposed method determines valid confidence bounds. Results are shown for smaller effect sizes ($\theta\leq 1.2$) and the extreme values of the denseness of the signal ($m_1\in\{10,90\}$).

In particular, Figure \ref{figure:Comparison_theta_rho_03} displays results under dependency of the p-values ($\rho=0.3$) and correct specification of $F$.
Results both for $\gamma^*=0.8$ and $\gamma^*=0.95$ are shown, as the first corresponds to valid confidence bounds for very small effect sizes ($\theta\geq 0.6$), while the latter corresponds to valid confidence bounds for slightly larger effect sizes ($\theta\geq 0.8$). 

Results under misspecification of $F$ ($\hat{\theta}=\theta+0.1$) and independence of the p-values are displayed in Figure \ref{figure:Comparison_theta_mis_rho_0}. Critical vectors were chosen using $\gamma^*=1$.

Lastly, results under dependency of the p-values ($\rho=0.3$) and misspecification of $F$ ($\hat{\theta}=\theta+0.1$) are displayed in Figure \ref{figure:Comparison_theta_mis_rho_03}. Critical vectors were chosen using $\gamma^*=1$.

\begin{figure}[p]
    \centering
    \includegraphics[width=1\textwidth]{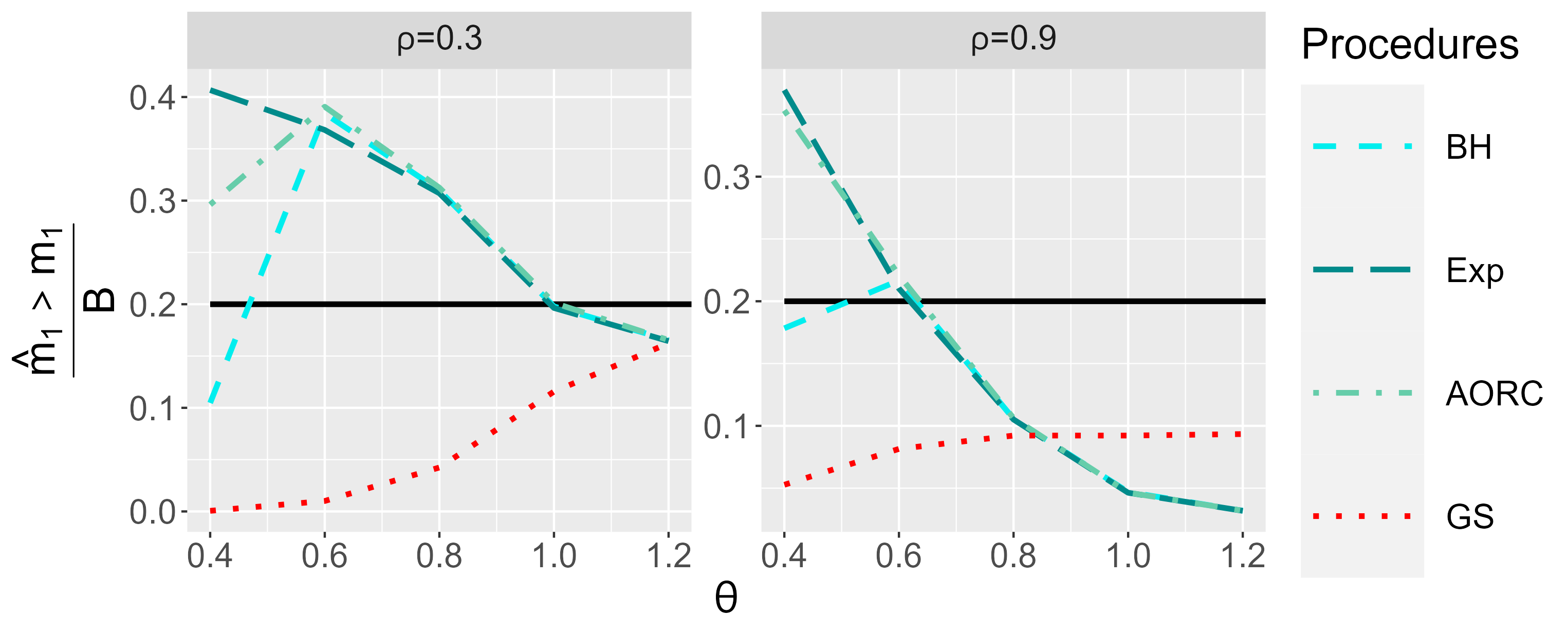}
   \caption{Dependency setting: p-values have correlation $\rho$ and effect size $\theta$ is known. Proportion of iterations for which $\hat{m_1}>m_1$, where $\hat{m_1}$ is the lower confidence bound for the number of true discoveries $m_1=50$ obtained from the proposed procedure with different critical vectors ($\BH$, $\Exp$, $\AORC$) and the method of \protect{\citet{GoemanSolari}}. The total number of iterations is $B=10{,}000$. The black solid line corresponds to the significance level $\alpha=0.2$.}
   \label{figure:Comparison_violation}
\end{figure}

Additionally, we have considered the setting with dependent p-values and correct specification of $F$ and we have studied the validity of the confidence bounds $\hmi$ obtained using any family with $\gamma^*=1$. Figure \ref{figure:Comparison_violation} illustrates the proportion of iterations for which the lower confidence bound is larger than the true number of discoveries
($m_1<\hat{m_1}$). If this proportion is larger than $\alpha$, the bounds are considered to be not valid.
Again, results are shown for smaller effect sizes ($\theta\leq 1.2$).

\begin{table}[ht]
\caption{Independence setting: p-values have correlation $\rho=0$ and the effect size $\theta$ is known. 
Sum of the expected values $\mathbb{E}_{\tmi}[R\gamma^*]$ over all possible candidate values $\tmi\in\{0,\ldots,m\}$, as given in Eq.~\eqref{eq:sum_exp}.
Results are obtained from the proposed procedure with different families of critical vectors ($\BH[\gamma^*]$, $\Exp[\gamma^*]$, $\AORC[\gamma^*]$). Bold values correspond to the highest value of the sum for each $\theta$.}
\label{table:theta_sum_exp}
\centering
\begin{tabular}{lrrrrr}
  \hline
   &\multicolumn{5}{c}{$\theta$}\\
  Critical vector & 0.6 & 0.8 & 1 & 1.2 &2 \\ 
  \hline
$\BH$ & 4720.33 & 4910.98 & 5011.96 & \textbf{5058.00} & 5065.01 \\ 
$\BH[0.95]$ & 4535.58 & 4702.08 & 4786.21 & 4817.20 & 4820.65 \\ 
$\BH[0.9]$ & 4372.23 & 4504.50 & 4569.98 & 4582.53 & 4584.15 \\
$\BH[0.8]$ & 3928.18 & 4079.47 & 4113.83 & 4110.73 & 4106.56 \\ 
\hline
$\BY$ & 4720.44 & 4911.31 & 5011.99 & 5057.98 & 5065.04 \\ 
$\BY[0.95]$ & 4540.14 & 4700.22 & 4786.94 & 4817.05 & 4820.16 \\ 
$\BY[0.9]$ & 4365.37 & 4505.53 & 4567.08 & 4583.50 & 4583.98 \\ 
$\BY[0.8]$ & 3921.54 & 4065.40 & 4111.10 & 4106.36 & 4111.25 \\ 
\hline
$\Exp$ & 4703.07 & 4909.88 & \textbf{5012.18} & 5057.61 & 5057.37 \\ 
$\Exp[0.95]$ & 4548.57 & 4686.55 & 4783.07 & 4818.85 & 4878.56 \\
$\Exp[0.9]$ & 4384.82 & 4481.32 & 4556.33 & 4589.78 & 4724.74 \\
$\Exp[0.8]$ & 3939.10 & 4030.66 & 4084.48 & 4119.19 & 4349.03 \\
\hline
$\AORC$ & \textbf{4740.10} & \textbf{4915.25} & 5012.17 & \textbf{5058.00} & \textbf{5065.06} \\ 
$\AORC[0.95]$ & 4604.67 & 4718.93 & 4786.76 & 4815.32 & 4820.62 \\ 
$\AORC[0.9]$ & 4455.21 & 4531.71 & 4388.23 & 4584.36 & 4586.82 \\ 
$\AORC[0.8]$ & 4011.63 & 4080.76 & 4101.44 & 4105.72 & 4106.77 \\ 
   \hline
\end{tabular}
\end{table}

\begin{table}[ht]
\caption{Independence setting: p-values have correlation $\rho=0$ and the effect size $\theta=0.8$ is known. Lower confidence bound $\hat{m}_1$ for the number of true discoveries $m_1$ obtained from the proposed procedure with different families of critical vectors ($\BH[\gamma^*]$, $\Exp[\gamma^*]$, $\AORC[\gamma^*]$) and the method of \citet{GoemanSolari} ($GS$). Bold values correspond to the highest value of $\hat{m}_1$ for each $m_1$.}
\label{table:theta_08_rho_0}
\centering
\begin{tabular}{lrrrrrrrrrr}
  \hline
 $m_1$&  10 & 20 & 30 & 40 & 50 & 60 & 70 & 80 & 90 \\ 
  \hline
$GS$ &  8.28 & 16.87 & 25.78 & 35.08 & 44.66 & 54.46 & 64.67 & 75.15 & 86.31 \\ 
  $\BH$ & 8.87 & 18.80 & 28.78 & 38.85 & 48.85 & 58.72 & 68.60 & 78.30 & 87.95 \\ 
  $\BY$ &  8.87 & 18.80 & 28.79 & 38.85 & 48.85 & 58.72 & 68.60 & 78.30 & 87.95 \\ 
  $\Exp$ &  9.01 & \textbf{18.92} & \textbf{28.86} & \textbf{38.88} & 48.83 & 58.66 & 68.50 & 78.18 & 87.82 \\ 
  $\AORC$ & 8.75 & 18.66 & 28.66 & 38.78 & \textbf{48.86} & \textbf{58.80} & \textbf{68.75} & \textbf{78.52} & \textbf{88.21} \\ 
  \hline
  $\BH[0.95]$  & 9.07 & 18.70 & 28.19 & 37.83 & 47.34 & 56.76 & 66.06 & 75.40 & 84.35 \\ 
  $\BY[0.95]$  & 9.06 & 18.69 & 28.17 & 37.81 & 47.31 & 56.74 & 66.04 & 75.38 & 84.33 \\ 
  $\Exp[0.95]$  & \textbf{9.13} & 18.69 & 28.10 & 37.70 & 47.12 & 56.55 & 65.78 & 75.16 & 84.07 \\ 
  $\AORC[0.95]$  & 9.03 & 18.69 & 28.22 & 37.91 & 47.52 & 56.96 & 66.38 & 75.68 & 84.71 \\ 
   \hline
\end{tabular}
\end{table}

\newpage
\clearpage
\section{Analysis of fMRI data}
\label{sec:Analysis_fMRI}
This section provides additional information on the analysis of fMRI data illustrated in Section \ref{sec:Case_Study}.
First, we describe in greater detail the workflow of the analysis. Then we describe how the ROIs for the considered data set have been defined.

\subsection*{Pipeline for the analysis}
The general process of the proposed fMRI analysis is illustrated in Figure \ref{figure:General_process}. The goal of the analysis is computing a lower ($1-\alpha$)-confidence bound for the proportion of active voxels (TDP) within a pre-specified ROI. The analysis uses the measured bold time series $\boldsymbol{Y}_{ij}$ and the design matrix $\boldsymbol{X}_{ij}$ for each voxel $i$ and each subject $j$. The distribution $F$ under the alternative of the p-values obtained from the second level analysis is assumed to be known, i.e., the parameters that characterize $F$ are assumed to be determined prior to the analysis.

\begin{figure}
\begin{tikzpicture}[scale=0.8]
    \draw[thick, fill=teal!15] (1.5,0) rectangle (15.5,4.75);
    \draw[thick, fill=teal!30] (1.5,5) rectangle (15.5,9);
    \draw[thick, fill=teal!5](0,0) rectangle (1.5,9);
    \node [align= center, rotate = 90] at (0.75,4.5) {Contrast map per subject (output first level analysis)\\+ Mask defining ROI};
   
    \node[align = left, right] at (1.5,8.5) {\underline{Subset with $N^{[e]}=50$ subjects}};
    \node[align = left, right] at (1.5,4.25){\underline{Subset with $N^{[b]}=90$ subjects}};
    
    \draw[->](1.5,7)--(3.5,7);
    \node[align=left, left, scale=0.85] at (3.5,7) { two-sided \\  one-sample \\  t-test; \\ $H_0:\mu_i=0$};
    \node[draw, align = left, right] at (3.6, 7){t-value map \\ p-value map};
    \draw[->](1.5,1)--(3.5,1);
    \node[align=left, left, scale=0.85] at (3.5,1) {two-sided \\  one-sample \\  t-test; \\ $H_0:\mu_i=0$};
    \node[draw, right] at (3.6, 1){p-value map};
   
    \draw[->](6,7)--(10.15,7);
    \node[align=left, right, scale=0.85] at (6,7){Select t-values $z$ based \\  on corresponding p-values};
   
    \node[draw,align= right, right] at (10.25,7){$\hat{\mu}_t^{[e]}$ as in Eq.~\eqref{eq:est_ef_size};\\
    $\nu^{[e]}=N^{[e]}-1$};
    \draw[->](11,6.25)--(11,5.75);
    \node[above] at (11,5.1){$\hat{\theta}=\hat{\mu}_t^{[e]}\cdot\sqrt{2/N^{[e]}}$};
    
    \draw[->](11,5.2)--(11,4.5);
    \node at (11,4.1){$\hat{\mu}_t^{[b]}=\hat{\theta}\cdot \sqrt{N^{[b]}/2}$; $\nu^{[b]}=N^{[b]}-1$};
   
    \draw[->](8.5,3.7)--(8.5,3);
    \node[align= left] at (7.75,2.5) {Define critical vector\\ $(t_1,\ldots,t_m)^\top$};
    \draw(8.5,2)--(8.5,1);
    \draw[<->](9.5,2.5)--(10.9,2.5);
    
     \draw[->](12.5,3.7)--(12.5,3);
    \node[draw] at (12.5,2.5) {$\gamma^*$ as in Eq.~\eqref{eq:hmi}};
    \draw(12.5,2)--(12.5,1);
    
    \draw[->](6,1)--(8.9,1);
    \node[align= left,right, scale=0.85] at (6.25,1){Apply \\  step-up test};
    \node[draw,align=left, right] at (9,1) {Observed number \\ of rejections $r$};
    \draw[->](12.3,1)--(12.9,1);
    \node[draw,right]at (13,1) {$\hat{m}_1$ as in Eq.~\eqref{eq:hmi}};
\end{tikzpicture}
\caption{Illustration of the work flow to compute a lower ($1-\alpha$)-confidence bound for the TDP within a given ROI. The sample is split into two sub-samples to estimate the effect size and compute the lower confidence bounds separately. For each sub-sample, the inputs are the contrasts $D_{ij}=\mu_i+\xi_{ij}$ 
   of each voxel $i$ and subject $j$.}
   \label{figure:Flow_chart}
\end{figure}
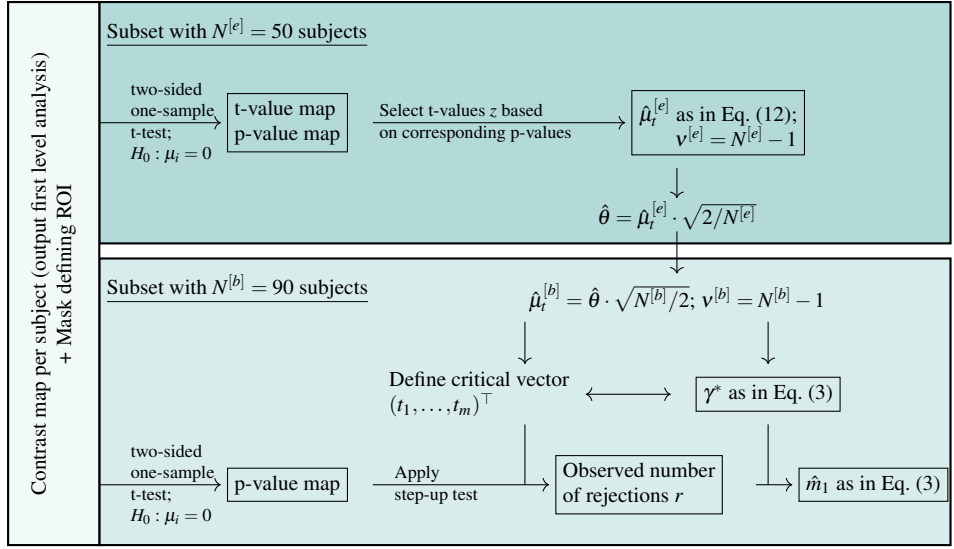

Figure \ref{figure:Flow_chart} further illustrates the last part of the process, where the contrasts obtained from the first level analysis are used to compute the TDP confidence bound. The $N=140$ subjects are split into two sub-groups; the first is used to estimate the effect size $\theta$, and the second to compute the confidence bound.

\begin{figure}
\begin{tikzpicture}[scale=0.75]
   
    \draw[fill=teal!5] (0,0) rectangle (3.9,7.5);
    \draw[fill=teal!10] (4,0) rectangle (7.9,7.5);
    \draw[fill=teal!20] (8,0) rectangle (11.9,7.5);
    \draw[fill=teal!30] (12,0) rectangle (15.9,7.5);
    
    \node[isosceles triangle,  fill=teal!50, isosceles triangle apex angle=90,
    minimum size =1.5cm, rotate=270] (T90) at (1.95,8.75){};
    \node[align=center, scale=0.9] at (1.95,8.75) {Observed BOLD\\$\boldsymbol{Y}_{i,j}$};
    \node[isosceles triangle,  fill=teal!50, isosceles triangle apex angle=90,
    minimum size =1.5cm, rotate=270] (T90) at (9.95,8.75){};
    \node[align=center, scale=0.9] at (9.95,8.75) {Mask defining\\ROI};
     \node[isosceles triangle,  fill=teal!50, isosceles triangle apex angle=90,
    minimum size =1.5cm, rotate=270] (T90) at (13.95,8.75){};
    \node[align=center, scale=0.9] at (13.95,8.75) {Parameters\\of $F$};
    
    \draw[-Triangle Cap,line width=5mm, teal!50] (0,7.5)--(4,7.5);
    \node[right] at (0,7.5){Preprocessing};
    \draw[-Triangle Cap,line width=5mm, teal!50] (4,7.5)--(8,7.5);
    \node[right] at (4,7.5){First Level Analysis};
    \draw[-Triangle Cap,line width=5mm, teal!50] (8,7.5)--(12,7.5);
    \node[right] at (8,7.5){Second Level Analysis};
    \draw[-Butt Cap,line width=5mm, teal!50] (12,7.5)--(15.9,7.5);
    \node[right] at (12,7.5){Inference on TDP};
    
    \node[scale=0.9] at (5.95,7) {Within-subject analysis};
    \node[scale=0.9] at (9.95,7) {Between-subject analysis};
    \node[scale=0.9] at (13.95,7) {Confidence bounds};
    
    \node[align=center] at (1.95,6.25){For each subject $j$,\\$j=1,\ldots,N$:};
    \node[] at (5.95,6.25){$\Tilde{\boldsymbol{Y}}_{ij}=\boldsymbol{X}_{ij}\cdot\boldsymbol{\beta}_{ij}+\epsilon_{ij}$};
    \node[] at (9.95,6.25){$D_{ij}=\mu_i+\xi_{ij}$};
  
    \node[below, align=left] at (1.95,5.4){$\cdot$ Slice Time Correction \\ $\cdot$ Registration\\ $\cdot$ Motion Correction\\ $\cdot$ Brain Extraction \\ $\cdot$ Spatial Smoothing \\ $\cdot$ High-pass Filtering};
    \node[below, align=left] at (5.95,5.48){$\cdot \boldsymbol{X}_{ij}\in R^{W\times K}$\\ \quad design matrix \\ $\cdot\boldsymbol{\beta}_{ij}:(\beta_{ij1},\ldots,\beta_{ijK})^\top$ \\ $\cdot1,\ldots,K$ conditions};
    \node[above, align=center] at (5.95, 2.2){Contrast:\\$D_{ij}=\hat{\beta}_{ij1}-\hat{\beta}_{ij2}$};
    \node[below, align=left] at (9.95,5.4){$\cdot\mu_i:$ difference in \\ \quad activation between \\ \quad conditions per voxel};
    \node[above, align=center] at (9.95, 2.2){One-sample, two-sided\\ t-test: $H_{0i}:\mu_i=0$};
    \node[below, align=left] at (13.95,5.4){$\cdot$ Define critical vector\\ $\cdot$ Compute $\gamma^*$ \\ $\cdot$ Determine $r$};
    \node[above, align=center] at (13.95, 2.2){Bounds:\\ $\hat{m}_1=\lceil r\cdot \gamma^* \rceil$};
  
    \node[below, align=center] at (1.95,2){\underline{Output}\\ Preprocessed BOLD \\$\Tilde{\boldsymbol{Y}}_{ij}$ for each voxel $i$\\ at time $w=1,\ldots W$};
    \node[below, align=center] at (5.95,2){\underline{Output}\\ Contrast $D_{ij}$ between \\ conditions 1 and 2};
    \node[below, align=center] at (9.95,2){\underline{Output}\\ p-value map\\ i.e., p-values per voxel \\ in the ROI};
   \node[below, align=center] at (13.95,2){\underline{Output}\\ Lower confidence \\ bounds of the TDP \\ for the ROI};
\end{tikzpicture}
 \caption{The general steps needed to compute a lower ($1-\alpha$)-confidence bound for the TDP within a given ROI. The inputs are, for each voxel and each subject, the measured BOLD time series $\boldsymbol{Y}_{ij}$ and the design matrix $\boldsymbol{X}_{ij}$.
   }
    \label{figure:General_process}
\end{figure}
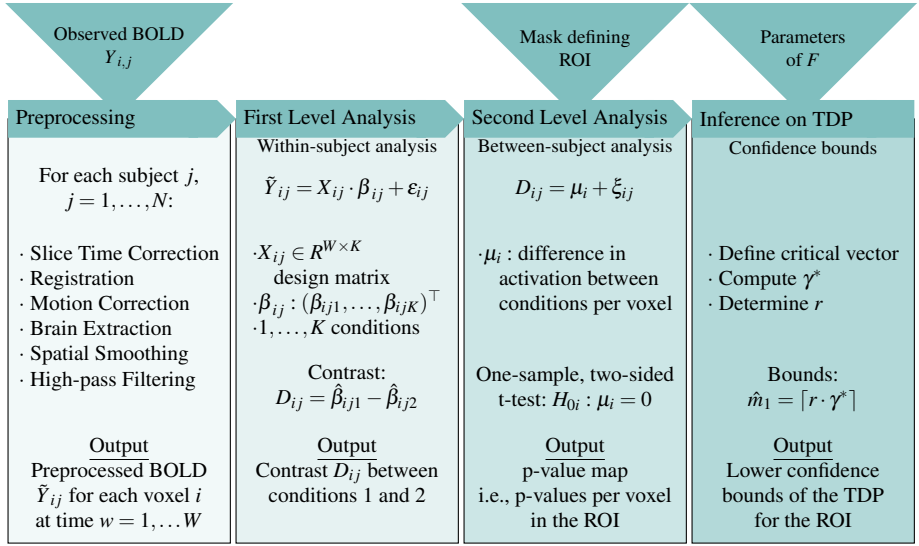

\subsection*{Definition of the ROI}
\label{sec:Def_ROI}
ROIs have been defined using the clusters and peaks of activation reported by \citet{Schirmer2012} and \citet{Binder2000}, that investigated human voices vs.~non-human sounds. In \citet{Schirmer2012}, we have considered only the four clusters containing at least 20 voxels with a total of seven peaks of activation. Additionally, we considered two clusters determined by \citet{Binder2000} with a total of eleven peaks of activation. Then we have defined the ROIs from the selected clusters, as following.

\begin{figure}[ht]
    \centering
    \includegraphics[width=1\textwidth]{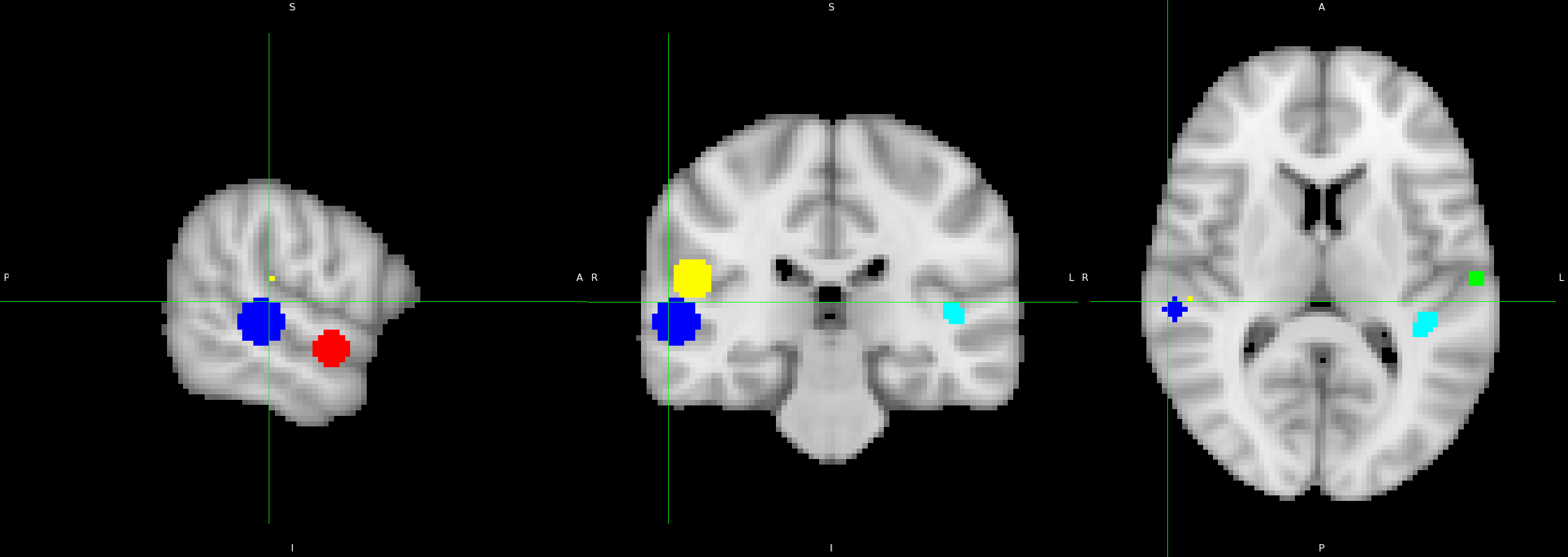}
   \caption{Location of the ROI in the brain. The turquoise region corresponds to the ROI in the Left STG, the green sphere corresponds to the ROI centered in the Left AC, the yellow sphere corresponds to the ROI centered in the Right AC, the blue sphere corresponds to the ROI centered in the Right MTG and the red sphere corresponds to the ROI centered in the Right STG. Note that in the picture on the left the right hemisphere of the brain is shown.}
   \label{figure:ROI_Schirmer_left}
\end{figure}

ROIs have been defined as either spheres or cuboids. Spherical ROIs have been defined such that they include all peaks of activation in the cluster and have a similar size in $mm^3$ as the clusters reported in the corresponding study. Note that only \citet{Schirmer2012} determined the size in $mm^3$ of the clusters. 
If the cluster contained more than one peak of activation, the center of the sphere has been defined as the mean voxel coordinate; if the cluster contained only one peak of activation the spherical ROI has been centered around it. If the spherical cluster had a voxel size that differed too much from the one given in \citet{Schirmer2012}, a cuboid ROI has been defined, such that each peak of activation was the center of a cube and these cubes were connected. The size of the cubes has been defined such that the resulting size of the ROI was similar to the cluster size given in \citet{Schirmer2012}. An overview of the different clusters, including their size and location, is given in Table \ref{table:Clusters_auditory} and illustrated in Figure \ref{figure:ROI_Schirmer_left}.

We have used the ``icbm2tal'' transformation in GingerAle (\citet{Laird2010}, \citet{Lancaster2007}, see also \url{https://www.brainmap.org/icbm2tal/}) to transform the Talairach coordinates given in \citet{Schirmer2012} to MNI coordinates. We have used FSL \citep{FSL} to create the masks for the ROIs.

\begin{table}[ht]
\caption{Information about the regions of interest. The clusters are named according to the area of the brain in which the peak of activation is located, based on the Talairach Daemon Labels in FSL \citep{Lancaster2000}. Given is the shape of the ROI, the width (for cuboid) or radius (for spheres) of the ROIs, the number of peaks of activation and the size of the ROI in terms of number of voxels. Furthermore, for spherical ROI the voxel coordinates [x,y,z] of the center of each sphere are reported, for the cuboid ROI the coordinates of the peaks of activation are reported. The coordinates are given in the MNI152 space.}
\label{table:Clusters_auditory}
\centering
\begin{tabular}{llllllrrr}
  \hline
Reference &ROI & Shape& Radius/width & Size &|Peaks| &  \multicolumn{3}{r}{Coordinates}    \\ 
&& & in mm& & &x&y&z \\
  \hline
\citet{Schirmer2012} & L STG &Cuboid&6 & 162&3& -38& -38& 10 \\
& & & &&&-44&-30&8\\
& & & &&&-50&-20&2\\
&L AC &Sphere& 4 &33 &2&-60&-18&12\\
&R AC &Sphere&8&257&1 &52& -26& 18 \\
&R FG &Sphere & 5  &81&1  &39&-39&-14\\
\citet{Binder2000}&R MTG & Sphere & 8 & 257&6&64&-4&-8\\
&R STG & Sphere & 9 &389 &5 &58&-30&2\\
   \hline
\end{tabular}
\end{table}

\end{document}